\documentclass[lettersize,journal]{IEEEtran}
\usepackage{amsmath,amsfonts}
\usepackage{amssymb}
\usepackage{algorithmic}
\usepackage{algorithm}
\usepackage{array}
\usepackage[caption=false,font=footnotesize,labelfont=rm,textfont=rm]{subfig}
\usepackage{textcomp}
\usepackage{stfloats}
\usepackage{subfig}
\usepackage{url}
\usepackage{verbatim}
\usepackage{graphicx}
\usepackage{cite}
\usepackage{hyperref}
\usepackage{tabularray}
\usepackage{color}
\usepackage{booktabs}
\newtheorem{theorem}{Theorem} 
\newtheorem{definition}{Definition} 
 
\begin{document}

\title{Ray Antenna Array Achieves Uniform Angular Resolution Cost-Effectively for Low-Altitude UAV Swarm ISAC}

\author{Haoyu Jiang and Yong Zeng,~\IEEEmembership{Fellow,~IEEE}
\thanks{ H. Jiang and Y. Zeng are with the National Mobile Communications Research Laboratory, Southeast University, Nanjing 210096, China. Y. Zeng is also with the Purple Mountain Laboratories, Nanjing 211111, China (e-mail: \{213211563, yong\_zeng\}@seu.edu.cn). (Corresponding author: Yong Zeng.)}}



\maketitle

\begin{abstract}
Ray antenna array (RAA) is a novel multi-antenna architecture comprising massive low-cost antenna elements and a few radio-frequency (RF) chains. The antenna elements are arranged in a novel ray-like structure, where each ray corresponds to a simple uniform linear array (sULA) with deliberately designed orientation and all its antenna elements are directly connected. 
By further designing a ray selection network (RSN), appropriate sULAs are selected to connect to the RF chains for further baseband processing. Compared to the classic hybrid analog/digital beamforming based on conventional antenna arrays, RAA has three appealing advantages: (i) dramatically reduced hardware cost since no phase shifters are needed; (ii) enhanced beamforming gain as antenna elements with higher directivity  can be used; (iii) uniform angular resolution across all signal directions. Such benefits make RAA especially appealing for integrated sensing and communication (ISAC), particularly for low-altitude unmanned aerial vehicle (UAV) swarm ISAC, where high-mobility aerial targets may easily move away from the boresight of conventional antenna arrays, causing severe communication and sensing performance degradation. Therefore, this paper studies RAA-based ISAC for low-altitude UAV swarm systems. First, we establish an input-output mathematical model for RAA-based UAV ISAC and rigorously show that RAA achieves  uniform angular resolution for all directions through beam pattern analysis. Besides, we design the RAA orientation and RSN to fully reap its advantages. Furthermore, RAA-based ISAC with orthogonal frequency division multiplexing (OFDM) for UAV swarm is studied, and efficient algorithm is proposed for sensing target parameter estimation. Extensive simulation results are provided to demonstrate the significant performance improvement by RAA system over the conventional antenna arrays, in terms of sensing angular resolution and communication spectral efficiency, highlighting the great potential of the novel RAA system to meet the growing demands of low-altitude UAV ISAC.

\end{abstract}

\begin{IEEEkeywords}
ray antenna array (RAA), low-altitude UAV, ISAC, UAV swarm, uniform angular resolution.

\end{IEEEkeywords}

\section{Introduction}
\IEEEPARstart{L}{ow-altitude} unmanned aerial vehicles (UAVs) are regarded as important aerial platforms for the emerging low-altitude economy, driven by their flexible maneuverability, rapid deployment capabilities, and cost-effectiveness\cite{8918497}. The advent of UAV swarm technology has further amplified these advantages, enabling collaborative task execution through distributed sensing and autonomous coordination \cite{9372970,xu2025integratedsuperresolutionsensingsymbiotic}. Modern multi-functional UAV systems now support critical applications ranging from traffic congestion monitoring to disaster zone search-and-rescue operations, delivering unprecedented spatial coverage and real-time observational precision \cite{9598918,9592698}. However, the very attributes that make UAVs indispensable – affordability, ease of operation, and payload capacity\cite{9915359} – also render them vulnerable to malicious exploitation, including illegal surveillance, airspace trespassing, and contraband delivery by maliciously manipulated commercial UAVs \cite{9714798,9785379}. This duality underscores an urgent need for effective solutions that simultaneously enhance UAV operational capabilities while enforcing stringent airspace supervision. Such a challenge perfectly aligns with the core objectives of integrated sensing and communication (ISAC) systems \cite{9456851,dai2025tutorialmimoofdmisacfarfield}. ISAC has emerged as a new paradigm for next-generation wireless systems, enabling the co-design of communication and sensing functionalities through shared hardware infrastructure, signal processing algorithms, and spectral resources \cite{9724187,9606831,9705498,9737357}. 
This integration is particularly critical for low-altitude UAV swarm management, where real-time detection and reliable command transmission must coexist to prevent collisions and ensure mission success \cite{10129074,10104142,song2024overviewcellularisaclowaltitude}.

Multi-antenna or multiple-input multiple-output (MIMO) has been a key technology for both wireless communication and radar sensing for decades. By exploiting the additional spatial multiplexing, diversity and beamforming gains offered by large array apertures\cite{2025arXiv250116610C}, MIMO systems may simultaneously enhance communication and sensing performance. These advantages scale up with array size, driving modern systems toward massive MIMO deployments with dozens or even hundreds of antenna elements. 
At higher frequency regimes like millimeter-wave (mmWave) and Terahertz bands \cite{8732419}, the further scaling of array size creates additional  implementation challenges. While smaller wavelengths at higher frequencies allow denser antenna packing, they also exacerbate path loss and phase noise sensitivity. 
In addition, the conventional fully digital architecture, where each antenna element requires one dedicated radio frequency (RF) chain, incurs prohibitive hardware cost, signal processing complexity, and energy consumption \cite{10430216}. This becomes particularly acute in high-frequency systems where wideband operation demands ultra-fast analog-to-digital converters (ADCs) and precise phase synchronization across hundreds or even thousands of array elements. This issue becomes even more severe for 6G and beyond in the era of extremely large-scale MIMO (XL-MIMO) \cite{10496996}. Such prohibitive overheads necessitate innovative antenna architectures that preserve performance while radically reducing implementation difficulty.

To address such issues, researchers have been exploring various solutions, including analog beamforming\cite{1367565}, hybrid analog/digital beamforming (HBF)\cite{6717211,7913599}, lens antenna arrays\cite{6824769,7416205}, fluid antenna/movable antenna\cite{https://doi.org/10.1049/el.2020.2788,10286328,dong2024movableantennawirelesscommunicationsprototyping,10906511}, pinching antenna\cite{yang2025pinchingantennasprinciplesapplications} and tri-hybrid MIMO\cite{castellanos2025embracingreconfigurableantennastrihybrid}. 
With analog beamforming, phase shifters are applied in the analog domain to achieve directional signal reception/transmission, and only one RF chain is needed regardless of the number of antenna elements\cite{1367565,5783993,5447703}. However, analog beamforming is inherently limited to single-stream MIMO transmission. 
HBF further integrated analog and digital beamforming together to make multi-stream transmission possible and achieves comparable performance with respect to fully digital solution when the number of transmitted data stream is relatively small \cite{6717211,7913599,7880698,7445130}. However, in high-frequency systems such as mmWave and THz systems, accurate phase shifters are hard to design. 
On the other hand, lens antenna arrays can transform the signal from the antenna space to the beamspace that has much lower dimensions, so as to reduce the number of RF chains significantly \cite{7416205,6824769}. 
More recently, fluid or movable antenna systems have been proposed by dynamically optimizing antenna shape/positions at transceivers (Tx/Rx) \cite{10286328,dong2024movableantennawirelesscommunicationsprototyping}. Compared to fixed antennas, fluid/movable antenna systems achieve superior performance with equal or fewer antennas and RF chains \cite{10906511}. 
Besides, pinching antenna \cite{yang2025pinchingantennasprinciplesapplications} applies small dielectric particles to waveguides to build line of sight (LoS) paths for users.
Moreover, tri-hybrid MIMO architecture \cite{castellanos2025embracingreconfigurableantennastrihybrid} has been proposed to incorporate reconfigurable antennas with both digital and analog precoding. 
However, all the aforementioned methods require additional hardware cost such as massive number of phase shifters, bulky 
electromagnetic lenses, motors, long waveguides or reconfigurable antennas, resulting in extra energy consumption, implementation complexity, and/or responding time. 

To practically enable flexible beamforming for high-frequency systems like mmWave and THz systems, and further enhance wireless communication and sensing performance cost-effectively, recently a novel multi-antenna architecture termed ray antenna array (RAA) was proposed \cite{1}. RAA leverages the design degree of freedom (DoF) in the spatial domain by deliberately placing massive inexpensive antenna elements in a ray-like structure, where each ray corresponds to a so-called simple uniform linear array (sULA), for the fact that all the antenna elements within each ray are directly connected. This configuration enables each sULA to form a beam with the mainlobe direction matching the ray orientation without relying on any analog or digital beamforming. By arranging them in a ray-like pattern, for any desired signal direction, a beam can be formed by selecting the appropriate sULA, and the effective projected aperture remains almost the same across all directions. Furthermore, since each sULA is only responsible for a small angular range, antenna elements with directional beam pattern can be employed to enhance its beamforming gain. To further reduce hardware cost, a ray selection network (RSN) is incorporated to reduce the number of RF-chains. The RSN adaptively selects the proper sULAs to be connected to the limited RF-chains for further baseband processing. Thus, compared to the classic fully digital or hybrid beamforming based ULA, the RAA system can achieve even better beamforming capability and uniform angular resolution for all signal directions, while greatly reducing hardware cost by replacing expensive phase shifters with much cheaper antenna elements. Note that the price paid by RAA is its larger size to accommodate the more antenna elements. Fortunately, such an issue is greatly alleviated for high-frequency systems like mmWave or THz systems where the signal wavelengths are very small.

Motivated by the appealing advantages of RAA mentioned above and to fully reap its uniform angular resolution characteristic, in this paper we propose to use RAA for low-altitude UAV swarm ISAC systems. The main contributions of this paper are summarized as follows:
\begin{itemize}
    \item First, we investigate the novel RAA architecture for low-altitude UAV swarm  ISAC systems. The input-output communication and sensing signal models are derived for the RAA-based UAV ISAC system. Besides, by analyzing the beam pattern of RAA, we rigorously show that RAA can achieve uniform angular resolution across all signal directions, which is strictly better than the conventional ULA that achieves equal array gain. In particular, different from ULA that suffers from poor angular resolution when the user/target is far away from the array boresight, RAA can maintain its high resolution ability in all directions.
    Besides, we also show that with the same array gain, since each sULA is only responsible for a smaller portion of the angular range, RAA can achieve higher beamforming gain than the conventional ULA since it can use antenna elements with higher directivity, benefiting both communication and sensing.

    \item Next, based on the beam pattern analysis, we design the orientation of each sULA and the RSN to fully reap the advantages of RAA. We analyze the signal model and propose efficient sensing algorithms for RAA based UAV swarm ISAC with orthogonal frequency division multiplexing (OFDM). In particular, as RAA has different equivalent array response vectors from the conventional ULAs, traditional array structure based sensing algorithms like Periodogram cannot be directly applied. Therefore, multiple signal classification (MUSIC) algorithm tailored for RAA is proposed to estimate the AoAs of targets. Based on these results, we further transform the multiple target estimation problem into single target sensing problem by performing zero forcing (ZF) spatial beamforming on the received signal to filter out the sensing matrix of other targets. Then, for delay and Doppler estimation of each target, we exploit the frequency and symbol diversity of OFDM and apply the 2-D Periodogram algorithm to obtain the delay and Doppler pairs.

    \item We demonstrate the effectiveness of the proposed RAA based ISAC via extensive numerical results. In particular, we simulate a UAV swarm moving in the air with varying AoAs. For sensing, compared with classic discrete Fourier transform (DFT) codebook based hybrid analog/digital beamforming with ULA, the RAA systems can achieve robust angular estimation across all directions, while the performance of ULA degrades significantly as the targets moving away from the boresight, which confirms the analytical results. In extreme cases when the targets are along the extended line of the linear array, ULA is unable to sense any target even if super-resolution algorithm is used. Moreover, the proposed algorithm also shows effective performance in delay and Doppler estimation. For communication, RAA also outperforms conventional ULA in terms of communication rate, thanks to its larger beamforming gain enabled by antenna elements with higher directivity.  

\end{itemize}

\begin{figure*}[t]
\centering
\includegraphics[width=0.8\linewidth]{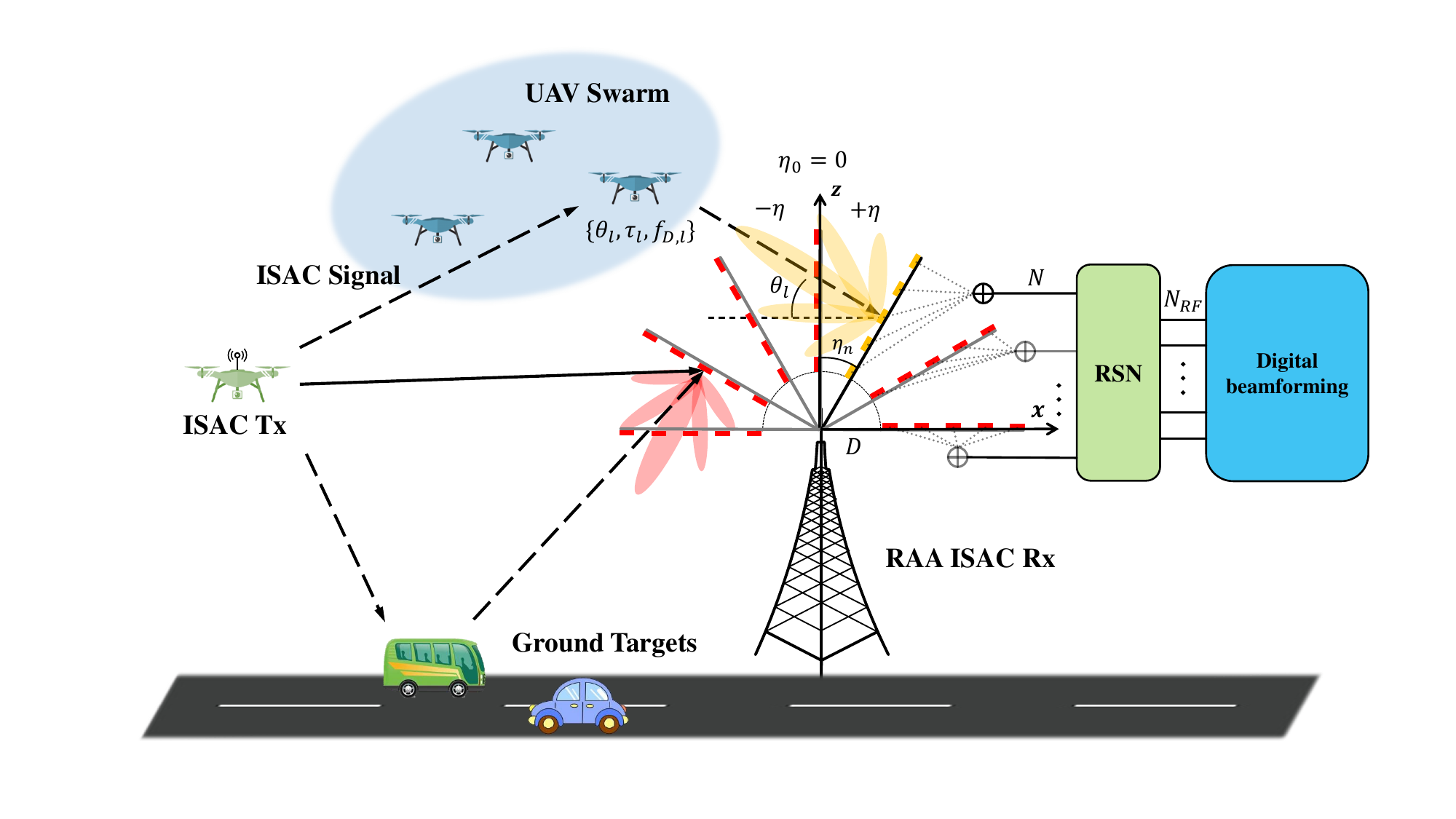}
\caption{An illustration of RAA-based ISAC for low-altitude UAV swarm.}
\label{fig:RAA env}
\end{figure*}

The rest of this paper is organized as follows. Section \ref{sec:system model} introduces system model for RAA based UAV swarm ISAC. The RAA architecture and equivalent channel model is presented in Section \ref{sec:RAA model}. After that, the comparison of RAA versus conventional ULA is discussed in Section \ref{sec:comparison}. In Section \ref{sec:algorithm}, the proposed sensing algorithm for RAA-based OFDM ISAC is presented. Simulation results and performance analysis are detailed in Section \ref{sec:simulation}. Finally, we conclude the paper in Section \ref{sec:conclusion}.

\textit{Notation}: Scalars are denoted by italic letters. Vectors and matrices are denoted by boldface lower and upper case letters, respectively.
The matrix inverse, transpose, and Hermitian transpose operations are given by $(\cdot)^{-1}$, $(\cdot)^{\mathrm{T}}$ and $(\cdot)^{\mathrm{H}}$, respectively.  The absolute value and $l_2$ norm are given by $|\cdot|$  and $\|\cdot\|_2$, respectively. $\mathbb{C}^{M\times N}$ denotes the space of $M \times N$ complex-valued matrices. The operator $\left \lceil \cdot \right \rceil $ is an integer ceiling operation, $\delta(\cdot)$ denotes the Kronecker delta function, and $\mathrm{card}(\cdot)$ denotes the cardinality of a set. $j = \sqrt{-1}$ denotes the imaginary unit of complex numbers, and $\mathbb{E}(\cdot)$ denotes the statistical expectation. The distribution of a circularly symmetric complex Gaussian (CSCG) random variable with zero mean and variance $\sigma^2$ is denoted by $\mathcal{CN}(0,\sigma^2)$ and $\sim $ stands for “distributed as”.




\section{System Model}\label{sec:system model}



As shown in Fig. \ref{fig:RAA env}, we consider a low-altitude UAV swarm ISAC system, which consists of a single-antenna ISAC Tx and a multi-antenna ISAC Rx. The channel from the Tx to the Rx comprises one LoS component and $L$ non-line-of-sight (NLoS) components induced by sensing targets. The $l$th path ($l=0$ for LoS) is parameterized by the tuple $\{ \theta_l,\tau_l,f_{D,l} \}$, which represents the AoA, path delay and Doppler frequency, respectively. The ISAC system simultaneously pursues three key objectives: (1) localization of the ISAC Tx UAV, (2) uplink communication from the ISAC-Tx to the BS, and (3) bi-static sensing of surrounding targets using reflected signals. 

To reduce the hardware cost while enhancing the performance, RAA is used at the ISAC Rx. The RAA architecture is composed by $N$ sULAs \cite{1}, making them ray-like as shown in Fig. \ref{fig:RAA env}. Each sULA has $M$ antenna elements separated by a distance $d=\lambda/2$, with $\lambda$ being the signal wavelength. Different from the conventional ULA, all the $M$ antenna elements of each sULA are directly combined without any analog or digital beamforming, which justifies the term sULA.
In addition, with only $N_{\mathrm{RF}} \le N$ RF chains, a RSN is designed to select $N_{\mathrm{RF}}$ ports from the $N$ equivalent sULA outputs. The RSN is denoted by $\mathbf{S} \in \{ 0,1 \}^{N_{\mathrm{RF}}\times N}$. Note that RSN has the function of selecting $N_{\mathrm{RF}}$ terms from $N$ elements, so the rows of $\mathbf{S}$ are composed of $N_{\mathrm{RF}}$ rows specially selected from an identity matrix $\mathbf{I}_{N}\in \mathbb{R}^{N\times N}$, i.e., $\mathbf{S} =[\mathbf{e}_{i_1},\mathbf{e}_{i_2},...,\mathbf{e}_{i_{N_{\mathrm{RF}}}}]^\mathrm{T}$, where $\mathbf{e}_{i_k}\in\mathbb{R}^{N\times 1},k=1,...,N_{\mathrm{RF}}$ denotes the $i_k$th row of $\mathbf{I}_{N}$. After that, the $N_{\mathrm{RF}}$ selected sULA outputs will be connected to the RF-chains for subsequent baseband digital processing for communication data decoding and sensing parameter estimation.

Let the ISAC signal transmitted by the Tx be $x(t)$.
The resulting signal at the $N$ sULAs of the RX, denoted by $\mathbf{\tilde{y}}(t)\in\mathbb{C}^{N\times 1}$, is expressed as
\begin{equation}\label{eq:rx antenna 1}
\begin{aligned}
    \mathbf{\tilde{y}}(t)&=\int \mathbf{h}(t,\tau)x(t-\tau)d\tau+\mathbf{z}(t),\\
\end{aligned}
\end{equation}
where $\mathbf{h}(t,\tau)\in\mathbb{C}^{N\times 1}$ is the equivalent channel between the Tx and the $N$ sULAs that will be modelled in Section \ref{sec:RAA model}, $\mathbf{z}(t)\in\mathbb{C}^{N\times 1}$ is the additive white Gaussian noise (AWGN) vector. After passing through the RSN $\mathbf{S}$, the resulting signal $\mathbf{y}(t)\in \mathbb{C}^{N_{\mathrm{RF}} \times 1}$ can be expressed as
\begin{equation}\label{eq:rx antenna 2}
\begin{aligned}
    \mathbf{y}(t)&=\mathbf{S} \mathbf{\tilde{y}}(t)\\&=\mathbf{S}\int \mathbf{h}(t,\tau)x(t-\tau)d\tau+\mathbf{S}\mathbf{z}(t).\\
\end{aligned}
\end{equation}

In the following, we will introduce the specific design of RAA, RSN, and derive the equivalent channel vector $\mathbf{h}(t,\tau)$.


\section{Ray Antenna Array}\label{sec:RAA model}

As a novel multi-antenna architecture, the RAA system employs $N$ radially arranged sULAs with $M$ elements each, as shown in Fig. \ref{fig:RAA env}. 
We establish a Cartesian coordinate system so that the sULA indexed by $n = 0$ aligns with the positive z-axis, as shown in Fig. \ref{fig:RAA env}. Furthermore, the orientation of the $n$th sULA with respect to (w.r.t.) the positive z-axis is denoted by $\eta_n \in [-\eta_{\max},\eta_{\max}]$, with $\eta_{\max}$ denoting the maximum orientation and $\eta_0 = 0$, where $n \in \mathcal{N}$ with $\mathcal{N} =\left \{ -\frac{N-1}{2},...,0,..., \frac{N-1}{2} \right \} $ being the set of sULA index, assuming $N$ is an odd number for notational convenience. To avoid strong mutual coupling for antenna elements across different sULAs, the first element of each sULA has a distance $D$ from the origin. Thus, the RAA is characterized by the set of parameters $\left ( N,M,D,\{ \eta_n \}_{n\in \mathcal{N}} \right )$. 

Consider the incoming signal from a certain path with a given elevation AoA w.r.t. the negative x-axis, denoted by $\theta$, with $\theta \in (-\pi/2,\pi/2]$.
Therefore the array response vector of a certain sULA with orientation $\eta$ for path angle $\theta$ can be expressed as \cite{1}
\begin{equation}\label{eq:RAA array response}
    \mathbf{a}(\theta,\eta)=[1,e^{j\pi \sin(\theta-\eta)},...,e^{j\pi(M-1) \sin(\theta-\eta)}]^\mathrm{T}.
\end{equation}

Thus, we may define the array response matrix of the RAA, denoted by $\mathbf{A}(\theta)\in\mathbb{C}^{M\times N}$, given by
\begin{equation}
    \mathbf{A}(\theta)=\big [\mathbf{a}(\theta,\eta_n)\big]_{n\in \mathcal{N}} \times \mathrm{diag}(\mathbf{b}),
\end{equation}
where $\mathbf{b}=\big [b(\theta-\eta_n)\big]_{n\in \mathcal{N}}\in \mathbb{C}^{N \times 1}$ captures the response of the reference element of each sULA, say the first element, given by
\begin{equation}
\label{eq:b}
    b(\theta-\eta_n)=e^{j\frac{2\pi}{\lambda}D\sin(\theta-\eta_n)}\sqrt{G(\theta-\eta_n)},
\end{equation}
where the term $e^{j\frac{2\pi}{\lambda}D\sin(\theta-\eta_n)}$ represents the phase shift of the first element of the $n$th sULA relative to the original point, and the second term $G(\theta-\eta_n)$ accounts the radiation pattern of each antenna element in the $n$th sULA, and it is characterized by the peak antenna gain $G(0)$, the 3dB beamwidth $\theta_{\mathrm{3dB}}$, and the total power gain denoted by $G_{\mathrm{sum}}=\int_{-\pi}^{\pi}G(\theta)d\theta$.

With all the $M$ antenna elements of each sULA directly connected, the resulted array response vector for all the $N$ sULAs, denoted by $\mathbf{r}(\theta) \in \mathbb{C}^{N\times 1}$, is expressed as:
\begin{equation}
\begin{aligned}
\label{eq:r}
    \mathbf{r}(\theta)&=\mathbf{A}(\theta)^\mathrm{T} \mathbf{1}_{M\times 1}=\mathrm{diag}(\mathbf{b})\Big[ \mathbf{a}(\theta,\eta_n)^\mathrm{T}\mathbf{1}_{M\times 1} \Big]_{n\in \mathcal{N}}\\
    &=\mathrm{diag}(\mathbf{b})\Big [MH_M\big(\sin(\theta-\eta_n)\big)\Big]_{n\in \mathcal{N}},
\end{aligned}
\end{equation}
where $H_M\big(\sin(\theta-\eta_n)\big)=\frac{1}{M}\mathbf{a}(\theta,\eta_n)^T\mathbf{1}_{M\times 1}$ is the Dirichlet kernel function given by:
\begin{equation}
\label{eq:Hm}
    H_M\big(x\big)=e^{j\frac{\pi}{2}(M-1)x}\frac{\sin\big({\frac{\pi}{2}Mx\big)}}{M\sin\big({\frac{\pi}{2}x\big)}}.
\end{equation}

Based on \eqref{eq:r}, the response of the $n$th sULA with AoA $\theta$ is
\begin{equation}\label{eq:rn}
    r(\theta,\eta_n)=Mb(\theta-\eta_n)H_M\big(\sin(\theta-\eta_n)\big).
\end{equation}

As a result, for the UAV swarm ISAC system with one LoS component and $L$ NLoS components, the equivalent channel $\mathbf{h}(t,\tau)$ in \eqref{eq:rx antenna 1} between the Tx and the $N$ sULAs of the RAA can be expressed as
\begin{equation}\label{eq:channel}
    \mathbf{h}(t,\tau)=\sum_{l=0}^L\alpha_l\mathbf{r}(\theta_l)\delta(\tau-\tau_l)e^{j2\pi f_{D,l}t},
\end{equation}
where $\alpha_l$ denotes the path coefficient, $\theta_l$,$\tau_l$, and $f_{D,l}$ respectively denote the path AoA, delay and Doppler frequency. By substituting \eqref{eq:channel} into \eqref{eq:rx antenna 2}, the resulting signal at the RX for RAA-based ISAC system is 
\begin{equation}\label{eq:rx antenna 3}
\begin{aligned}
    \mathbf{y}(t)=\mathbf{S}\sum_{l=0}^L\alpha_l\mathbf{r}(\theta_l)e^{j2\pi f_{D,l}t}x(t-\tau_l)+\mathbf{S}\mathbf{n}(t).\\
\end{aligned}
\end{equation}

Due to the limited RF chains, we need to choose $N_{\mathrm{RF}}$ out of $N$ RAA outputs. To fully exploit the energy-focusing ability of RAA, we propose the simple energy based ray selection, i.e., $\mathbf{S}$ is selected based on the this criterion:  $\max_{\mathbf{S}} \| \mathbf{y}(t) \|_2^2$, where the $N_{\mathrm{RF}}$ RAA outputs with the maximum sum energy are chosen.
To realize the energy based ray selection, we only need to sweep all the RAA ports to obtain full information about the response magnitude of different sULAs. This would require $\left \lceil \frac{N}{N_{\mathrm{RF}}} \right \rceil$ sweeps to cover all the N sULAs. Therefore, $\mathbf{S}$ will be determined and fixed for the subsequent analysis.

\section{RAA vs Conventional ULA}\label{sec:comparison}
On the basis of the RAA models presented in Section \ref{sec:RAA model}, in this section we analyze the corresponding beam patterns for RAA and compare it with conventional ULA which employs DFT codebook based HBF. Beam pattern characterizes the intensity variation of a beam designed for a certain desired direction $\theta'$ as a function of the actual observation direction $\theta$. It is crucial for ISAC as it reflects the level of angular resolution ability for sensing and inter-user interference (IUI) suppression capability for communication. For the conventional ULA with $M$ antenna elements, with a beamforming vector $\mathbf{v}(\theta')\in\mathbb{C}^{M\times 1}$ aiming to beamsteer energy towards the desired direction $\theta'$, the beam pattern at the observation direction $\theta$ is defined as:
\begin{equation}\label{eq:ULA beam pattern}
\begin{aligned}
    \mathcal{G}_{\mathrm{ULA}}(\theta,\theta')&\triangleq \Big|\sqrt{G_{\mathrm{ULA}}(\theta)}\mathbf{v}^\mathrm{H}(\theta')\mathbf{a}(\theta)\Big|\\
    &=\Big|\sqrt{G_{\mathrm{ULA}}(\theta)}\mathbf{a}^\mathrm{H}(\theta')\mathbf{a}(\theta)\Big|,
\end{aligned}
\end{equation}
where $\mathbf{a}(\theta)\in \mathbb{C}^{M\times 1}$ is the array response vector of ULA at the observation direction $\theta$ that will be given in this section. The second equality follows by applying the analog maximal-ratio transmission (MRT) beamforming, i.e. $\mathbf{v}(\theta')=\mathbf{a}(\theta')$, and $G_{\mathrm{ULA}}(\theta)$ denotes the radiation pattern of each
antenna element in ULA, characterized by the peak antenna gain $G_{\mathrm{ULA}}(0)$, the 3dB beamwidth $\theta_{\mathrm{3dB}}^{\mathrm{ULA}}$, and the total power gain denoted by $G^{\mathrm{sum}}_{\mathrm{ULA}}=\int_{-\pi}^{\pi}G_{\mathrm{ULA}}(\theta)d\theta$. For RAA, based on \eqref{eq:rn}, the desired direction $\theta'$ achieves the maximum power by selecting the sULA with array orientation matching the desired signal direction, i.e., $\eta_n=\theta'$. Therefore, the beam pattern of RAA can be defined as
\begin{equation}\label{eq:RAA beam pattern}
\begin{aligned}
    \mathcal{G}_{\mathrm{RAA}}(\theta,\theta')&\triangleq \Big|r(\theta, \theta')\Big|,
\end{aligned}
\end{equation}
where $r(\theta, \theta')$ is defined in \eqref{eq:rn}. Based on beam pattern, we can have the definition of angular resolution:

\begin{definition}\label{def:resolution}
     The angular resolution $\gamma$ of any type of array is a function of the desired signal direction $\theta'$, which is defined as half of the main lobe beam width $\Delta_\theta$ of its beam pattern $\mathcal{G}(\theta,\theta')$, i.e., 
\begin{equation}\label{eq:angular resolution RAA}
    \gamma(\theta')=\frac{1}{2}\Delta_\theta,
\end{equation}
\end{definition}
where $\Delta_\theta=|\theta_1-\theta_2|$, with $\theta_1$ and $\theta_2$ being the first right and left null points of $\mathcal{G}(\theta,\theta')$ for any given desired direction $\theta'$, i.e., 
\begin{equation}
    \theta_1 = \mathop{\min}\limits_{\theta}\ \Big\{\theta>\theta'\big|\mathcal{G}(\theta,\theta')=0\Big\}.
\end{equation}

\begin{equation}
    \theta_2 = \mathop{\max}\limits_{\theta}\ \Big\{\theta<\theta'\big|\mathcal{G}(\theta,\theta')=0\Big\}.
\end{equation}


\subsection{Beam Pattern and Angular Resolution of RAA}

\begin{figure*}[t]
  \centering
  \subfloat[RAA beam patterns\label{fig:RAA beam}]{
    \includegraphics[width=0.4\linewidth]{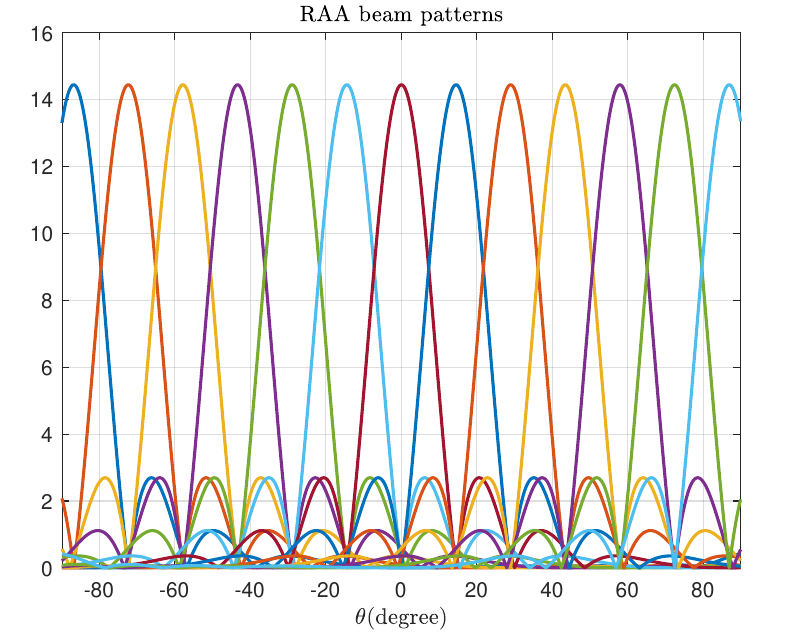}
  }
  \subfloat[HBF ULA beam patterns\label{fig:HBF beam}]
  {
    \includegraphics[width=0.4\linewidth]{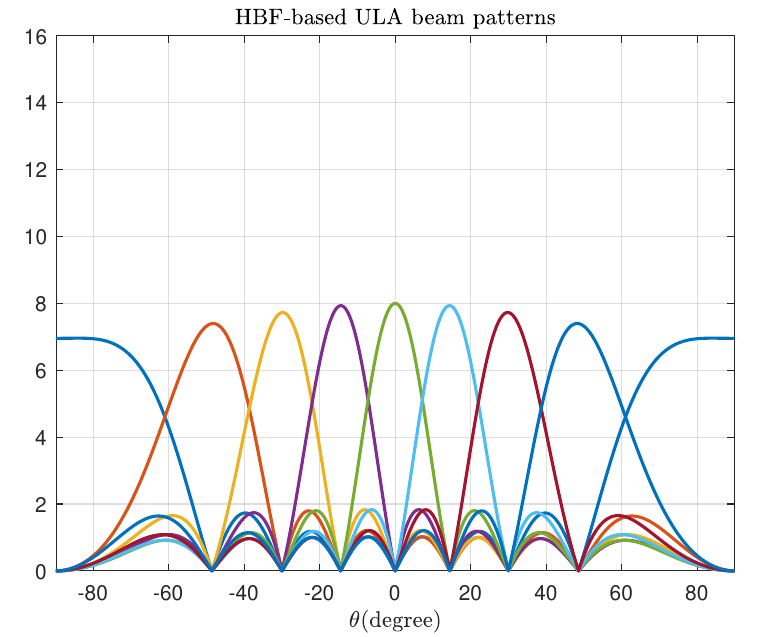}
  }
  \caption{The beam patterns for (a) RAA, where each curve represents the response of one sULA. (b) ULA, where each curve corresponds to one codeword in the DFT codebook. Both RAA and ULA achieves the same array gain $M=8$, but RAA achieves uniform angular resolution and higher beamforming gain.}
  \label{fig:beam pattern}
\end{figure*}

By substituting \eqref{eq:b} and \eqref{eq:rn} into \eqref{eq:RAA beam pattern}, the beam pattern of RAA can be expressed as 
\begin{equation}\label{eq:RAA beam pattern 2}
\begin{aligned}
    \mathcal{G}_{\mathrm{RAA}}(\theta,\theta')
    &=M\sqrt{G(\theta-\theta')}\bigg|H_M\big( \sin(\theta-\theta') \big)\bigg|.
\end{aligned}
\end{equation}

With beam pattern $\mathcal{G}(\theta,\theta')$ in \eqref{eq:RAA beam pattern 2}, we try to derive the angular resolution of RAA. For convenience, we make the assumption that the beamwidth of each antenna element $G(\zeta)$ is wider than that of $H_M\big( \sin(\theta-\theta') \big)$. Therefore, we have the following theorem:

\begin{theorem}\label{the:RAA resolution}
For any desired direction $\theta'$, the angular resolution of RAA is a constant, given by:
\begin{equation}
\begin{aligned}
\gamma_{\mathrm{RAA}}(\theta')=\gamma_{\mathrm{RAA}}=\arcsin\frac{2}{M}, \forall \theta'.
\end{aligned}
\end{equation}
\end{theorem}

\begin{IEEEproof}
Based on \eqref{eq:RAA beam pattern 2}, by letting $\sin(\theta-\theta')=\pm\frac{2}{M}$, we have $H_M\big(\sin(\theta-\theta')\big)=0$. Thus, according to definition \ref{def:resolution}, $\theta_1=\theta'+\arcsin(\frac{2}{M}),\theta_2=\theta'-\arcsin(\frac{2}{M})$, so the main lobe beam width can be obtained as $\Delta_\theta=|\theta_1-\theta_2|=2\arcsin(\frac{2}{M})$, which yields $\gamma_{\mathrm{RAA}}=\frac{1}{2}\Delta_\theta=\arcsin\frac{2}{M}$.

\end{IEEEproof}

Theorem \ref{the:RAA resolution} demonstrates that RAA achieves uniform angular resolution across all signal directions, i.e., it is independent of the desired signal direction $\theta'$. When $M\gg 1$, $\gamma_{\mathrm{RAA}}\approx \frac{2}{M}$, which is inverse proportional to the aperture of the sULA.
To achieve effective interference suppression between adjacent sULAs, as proposed in \cite{1}, the angular null position of one sULA's principal lobe must coincide with the peak direction of its adjacent counterpart. In this case, the orientation of the $N$ sULAs are designed as
\begin{equation}
\eta_n = n\times \arcsin(2/M),  \forall n \in \mathcal{N}.
\end{equation}

This closed-form solution ensures orthogonality between adjacent sULAs while maintaining full angular coverage capability.
There will be a total of $N=2\left \lfloor \frac{\eta_{\mathrm{max}}}{\arcsin(2/M)}+1 \right \rfloor $ sULAs in RAA. If the RAA needs to cover the half space so that $\eta_{\max}=\pi/2$, then $N\approx \lfloor(\pi M/2)\rfloor$ when $M\gg 1$. In addition, to ensure all antenna elements are separated by at least half wavelength, the distance $D$ needs to satisfy $D\ge\lambda/(4\sin(0.5\arcsin(2/M)))$\cite{1}. 

Note that for RAA, each sULA is only responsible for a small portion of the whole angular range from $[-\eta_{\mathrm{max}},\eta_{\mathrm{max}}]$. Therefore, different from the conventional antenna arrays where the antenna elements are shared by all beams or signal directions, we can use antenna elements with stronger directivity to enhance the overall beamforming gain as rigorously proved in \cite{RAAjournal}.

    



Fig. \ref{fig:RAA beam} illustrates the beam patterns of the different sULAs in the proposed RAA architecture, where $M=8$ and $\theta_{\mathrm{max}}=\eta_{\mathrm{max}}=\pi/2$. The antenna element pattern adopts the 3GPP radiation pattern \cite{3GPPchannel}, where we set $\theta_{\mathrm{3dB}}=0.3\pi$ and $G_{\mathrm{dB}}(0)=5.13$ dB.
It reveals two key characteristics of RAA:
\begin{itemize}
    \item Full angular coverage: For any signal with AoA $\theta$, there exists at least one  sULA exhibiting a dominant main lobe response.
    \item Uniform angular resolution: The resolution $\gamma_{\mathrm{RAA}}$ remains the same across all directions, as mathematically proved in Theorem \ref{the:RAA resolution} and verified numerically in Fig. \ref{fig:RAA beam}.
\end{itemize}


\subsection{Beam Pattern and Angular Resolution of Conventional ULA}
\begin{figure*}[t] 
        \centering \includegraphics[width=0.8\linewidth]{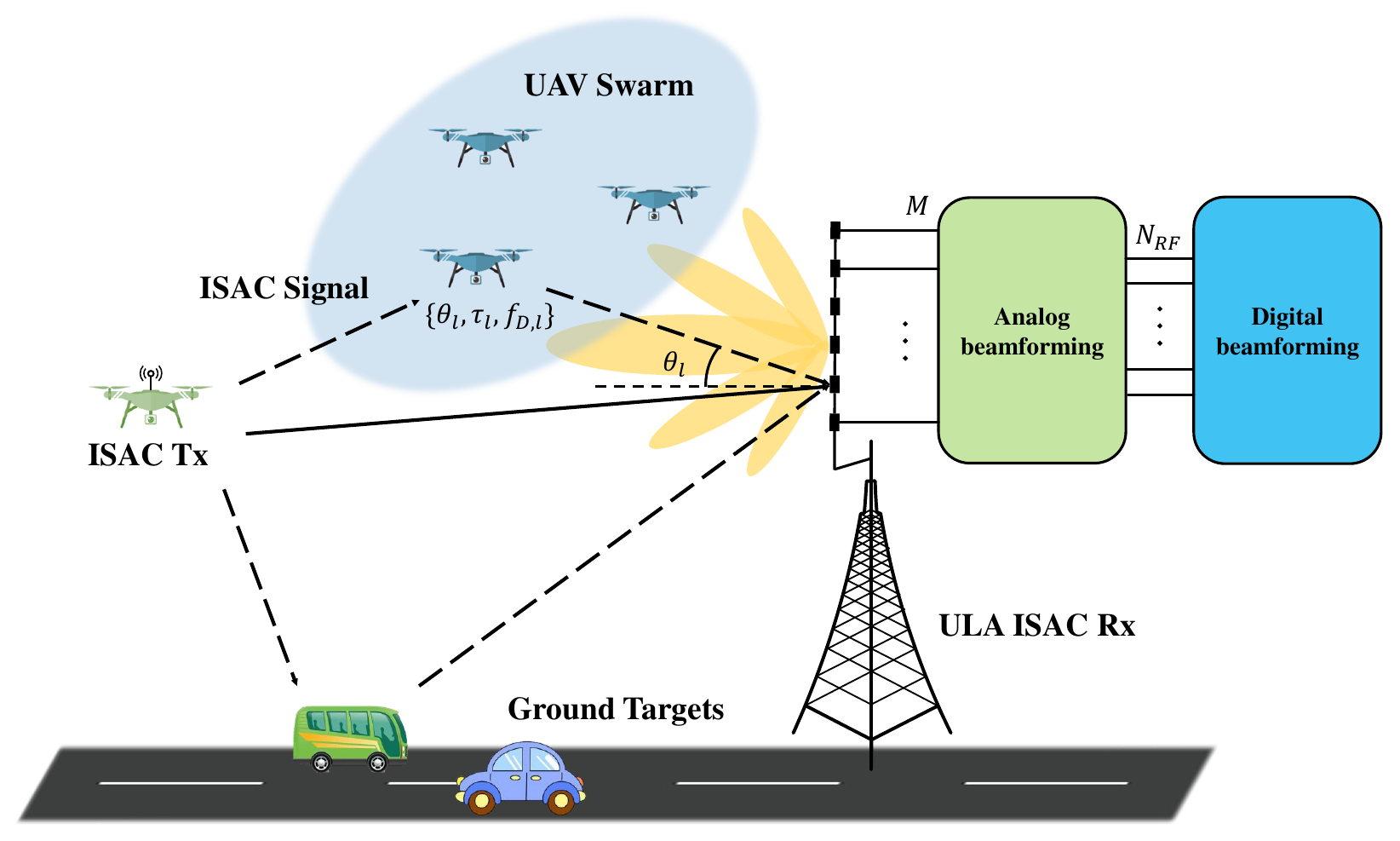}
        \caption{\label{fig:HBF env}An illustration of conventional ULA-based ISAC for low-altitude UAV swarm.}
\end{figure*}

As illustrated in Fig. \ref{fig:HBF env}, we use conventional ULA which employs DFT codebook based HBF as a benchmark comparison. 
To ensure fair comparison, we assume that the RAA and ULA have the same number of RF-chains $N_{\mathrm{RF}}$ and array gain $M$. The array response vector of the M-element
ULA is $\mathbf{a}(\theta)=[e^{j\pi (m-1)\sin\theta}]_{1\le m\le M}\in\mathbb{C}^{M\times 1}$. 
The DFT codebook for the HBF architecture is denoted by $\mathbf{A}_{\mathrm{DFT}}=\big[ \mathbf{a}(\varphi_{n}) \big]_{n\in\mathcal{N}'}\in\mathbb{C}^{M\times M}$, where $\mathbf{a}(\varphi_{n})=\big[ e^{j\pi(m-1)\sin\varphi_{n}} \big]_{1\le m\le M}\in\mathbb{C}^{M\times 1}$ is the $n$th DFT codeword, $\mathcal{N}'=\{1,2,...,M \}$ is the index set of DFT codewords, and $\sin\varphi_{n}=-1+2(n-1)/M$. 




Note that for ULA, the whole angular range needs to be covered by a single antenna array. This implies  that each element of ULA should have a wider coverage (or smaller directivity) than that of RAA, i.e. $\theta_{\mathrm{3dB}}^{\mathrm{ULA}}>\theta_{\mathrm{3dB}}$ \cite{RAAjournal}


By substituting the array response vector $\mathbf{a}(\theta)$ into \eqref{eq:ULA beam pattern}, the beam pattern of ULA is
\begin{equation}\label{eq:ULA beam pattern 2}
    \begin{aligned}
    \mathcal{G}_{\mathrm{ULA}}(\theta,\theta')
    &=M\sqrt{G_{\mathrm{ULA}}(\theta)}\Big|H_M\big( \sin\theta-\sin\theta' \big)\Big|.
\end{aligned}
\end{equation}

Note that at the first glance, the beam pattern of ULA in \eqref{eq:ULA beam pattern 2} looks very similar to that of RAA in \eqref{eq:RAA beam pattern 2}. However, they have two important differences. Firstly, they have different antenna element radiation pattern of $G_{\mathrm{ULA}}(\theta)$ and $G(\theta-\theta')$, respectively. Secondly, their variables in the Dirichlet kernel function are not the same, which are $\sin\theta-\sin\theta'$ and $\sin(\theta-\theta')$ for ULA and RAA, respectively.

Fig. \ref{fig:HBF beam} shows the beam patterns of ULA under different desired direction $\theta'$ for $M=8$ and $\theta_{\mathrm{max}}=\pi/2$ with $\theta_{\mathrm{3dB}}^{\mathrm{ULA}}=\pi$ and $G_{\mathrm{dB}}^{\mathrm{ULA}}(0)=0$dB,  ensuring that the total radiation power satisfies $G_{\mathrm{ULA}}^{\mathrm{sum}}=G_{\mathrm{sum}}$. 
It can be found that both RAA and ULA have full array gain when $\theta$ matches with $\theta'$, i.e. $H_M(\sin\theta-\sin\theta')\Big|_{\theta=\theta'}=1$ and $H_M(\sin(\theta-\theta'))\Big|_{\theta=\theta'}=1$ . However, with higher directional antenna elements, RAA can achieve higher overall beamforming gain compared to ULA. Besides, it is noteworthy that when $\theta$ grows larger, the mainlobe of ULA beam pattern becomes wider, as rigorously shown by the following theorem:

\begin{theorem}\label{the:HBF resolution}
The angular resolution of ULA as a function of the desired signal direction $\theta'$ is:
\begin{equation}
\begin{aligned}
\gamma_{\mathrm{ULA}}(\theta')=\frac{1}{2}\arcsin(\sin\theta'+\frac{2}{M})-\frac{1}{2}\arcsin(\sin\theta'-\frac{2}{M}),
\end{aligned}
\end{equation}
where $\theta'$ satisfies $-1+\frac{2}{M}\le\sin\theta'\le1-\frac{2}{M}$.
\end{theorem}

\begin{IEEEproof}
Based on Definition \ref{def:resolution} and $\mathcal{G}_{\mathrm{ULA}}(\theta,\theta')$ in \eqref{eq:ULA beam pattern 2}, by letting $\sin\theta-\sin\theta'=\pm\frac{2}{M}$, we will get $\theta=\arcsin(\sin\theta'\pm\frac{2}{M})$, and we have $H_M\big(\sin\theta-\sin\theta'\big)=0$. Therefore,  the main lobe beam width can be obtained as $\Delta_\theta=\arcsin(\sin\theta'+\frac{2}{M})-\arcsin(\sin\theta'-\frac{2}{M})$, which yields $\gamma_{\mathrm{ULA}}(\theta')=\frac{1}{2}\arcsin(\sin\theta'+\frac{2}{M})-\frac{1}{2}\arcsin(\sin\theta'-\frac{2}{M})$.


\end{IEEEproof}

Theorem \ref{the:HBF resolution} demonstrates that different from that for RAA in Theorem \ref{the:RAA resolution}, the angular resolution of ULA in general gets worse at higher AoA. 
In extreme cases where $\sin\theta'=\pm(1-\frac{2}{M})$, we have $\gamma_{\mathrm{ULA}}(\theta')=\frac{\pi}{4}-\frac{1}{2}\arcsin(1-\frac{4}{M})\ge\sqrt{\frac{2}{M}}$. When $M=16$, $\gamma_{\mathrm{ULA}}(\theta')\ge0.35\mathrm{rad}\approx20^\circ$.
Note that $\gamma_{\mathrm{ULA}}(0)=\arcsin(\frac{2}{M})=\gamma_{\mathrm{RAA}}$.
\begin{theorem}\label{the:resolution comparasion}
For equal array gain factor $M$, the angular resolution of RAA is always superior than that of ULA, i.e., 
\begin{equation}
\begin{aligned}
\gamma_{\mathrm{ULA}}(\theta')\ge\gamma_{\mathrm{RAA}}(\theta')=\arcsin(\frac{2}{M}), \forall\theta'. 
\end{aligned}
\end{equation}
where the equality holds if and only if $\theta'=0$.
\end{theorem}

\begin{IEEEproof}
Please refer to Appendix \ref{app:resolution}.


\end{IEEEproof}

\begin{figure}[t] 
        \centering \includegraphics[width=0.9\columnwidth]{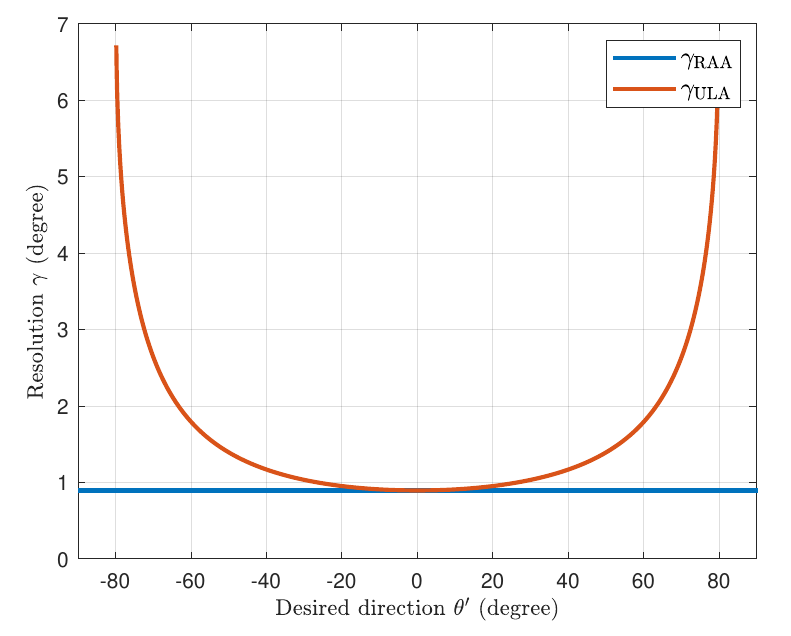}
        \caption{\label{fig:resolution compare}The comparision of angular resolution of RAA versus the conventional ULA, for $M=128$.}
\end{figure}

The resolution comparison established in Theorem \ref{the:resolution comparasion} is also illustrated in Fig. \ref{fig:resolution compare}. It is observed that when the desired direction $\theta'$ deviates from the boresight angle, the main lobe of the conventional ULA beam pattern becomes wider. When the communication user or sensing targets are located at a direction with large $\theta'$, ULA may fail to distinguish them due to the poor angular resolution. However, RAA is able to achieve uniform angular resolution of $\arcsin\frac{2}{M}$ across all directions $\theta'$, making it quite appealing for IUI suppression for multi-user communication and multi-target resolution for sensing. Besides, due to the fact that RAA does not use one single expensive phase shifter, it can significantly reduce the hardware cost. For example, in an example system with commerical TGP2108-SM 6-bit digital phase shifter and the QM12002 RF switch with $M=16$, the proposed
RAA can significantly reduce the hardware cost, amounting to less than 1\% of the HBF architecture \cite{1}.

In summary, the comparison of RAA and conventional ULA demonstrates the following advantages of RAA:
\begin{itemize}
    \item \textit{Higher beamforming gain}: 
    RAA has higher overall beamforming gain since it can use antenna elements with higher directivity.
    \item \textit{Uniform angular resolution}: 
    RAA can achieve uniform angular resolution across all signal directions.
    \item \textit{Cost effective}: 
    RAA is cost effective since no single phase shifter is needed, which is typically expensive and difficult to design, especially for high-frequency systems.
\end{itemize}

\section{Sensing Algorithm for RAA-based OFDM-ISAC}\label{sec:algorithm}
\subsection{RAA-based OFDM-ISAC}
In this section, we apply the proposed RAA-based ISAC system with OFDM waveform \cite{dai2025tutorialmimoofdmisacfarfield}.  Let $N_{\mathrm{sc}}$ denote the number of subcarriers and $M_{\mathrm{sym}}$ being the number of OFDM symbols within a coherent processing interval (CPI). The system bandwidth is $B$, and the subcarrier spacing is $\Delta f=B/N_{\mathrm{sc}}=1/T$, where $T$ represents the OFDM symbol duration before adding the cyclic prefix (CP). The duration of CP is denoted as $T_{\mathrm{cp}}$.  Then, the OFDM symbol duration including CP is $T_{\mathrm{s}}=T+T_{\mathrm{cp}}$. Denote the communication data of subcarrier $p$ and symbol $q$ as $d_{pq}\in \mathbb{C}$ with $\mathbb{E}[|d_{pq}|^2]=P_t$, where $P_t$ is the transmit power. After IFFT and adding CP, the transmitted signal $x(t)$ in \eqref{eq:rx antenna 1} can be expressed as \cite{dai2025tutorialmimoofdmisacfarfield}:
\begin{equation}\label{eq:x(t)}
    x(t)=\sum_{p=0}^{N_{\mathrm{sc}}-1}\sum_{q=0}^{M_{\mathrm{sym}}-1}d_{pq}e^{j2\pi p\Delta f(t-qT_{\mathrm{s}}-T_{\mathrm{cp}})}\mathrm{rect}\left ( \frac{t-qT_{\mathrm{s}}}{T_{\mathrm{s}}}\right ).
\end{equation}




Then, substituting \eqref{eq:channel} and \eqref{eq:x(t)} into \eqref{eq:rx antenna 1}, the resulting signal by the $N$ sULAs at the Rx side before RSN can be expressed as:
\begin{equation}
\begin{aligned}
\label{eq:y no rsn}
    \mathbf{\tilde{y}}(t)
    &=\sum_{l=0}^{L}\sum_{p=0}^{N_{\mathrm{sc}}-1}\sum_{q=0}^{M_{\mathrm{sym}}-1}\alpha_l\mathbf{r}(\theta_l)d_{pq}e^{j2\pi p\Delta f(t-\tau_l-qT_{\mathrm{s}}-T_{\mathrm{cp}})}\\
    &\times e^{j2\pi f_{D,l}t}\mathrm{rect}\left ( \frac{t-\tau_l-qT_{\mathrm{s}}}{T_{\mathrm{s}}}\right )+\mathbf{z}(t),
\end{aligned}
\end{equation}
where $\mathbf{z}(t)=\left[ \sum_{m=0}^{M-1}z_{m,n}(t) \right]_{n\in \mathcal{N}}$, and $z_{m,n}(t)$ denotes the noise from the $m$th element of the $n$th sULA.

By subsituting \eqref{eq:b} and \eqref{eq:r} into \eqref{eq:y no rsn}, the received signal of each of the $N$ sULA $\tilde{y}_n(t)$ can be further written as:
\begin{equation}
\begin{aligned}
\tilde{y}_n(t)&=\sum_{l=0}^{L}\sum_{p=0}^{N_{\mathrm{sc}}-1}\sum_{q=0}^{M_{\mathrm{sym}}-1}
\bar{\alpha}_{n,l}d_{pq}e^{j2\pi p\Delta f(t-\tau_l-qT_{\mathrm{s}}-T_{\mathrm{cp}})}\\
&e^{j2\pi f_{D,l}t}\mathrm{rect}\left ( \frac{t-\tau_l-qT_{\mathrm{s}}}{T_{\mathrm{s}}}\right )+z_n(t),
\end{aligned}
\end{equation}
where $\bar{\alpha}_{n,l}=\alpha_lb(\theta_l-\eta_n)H_M\left( \sin(\theta_l-\eta_n)  \right), \ z_n(t)=\sum_{m=0}^{M-1}z_{m,n}(t)$

By substituing \eqref{eq:channel} and \eqref{eq:x(t)} into \eqref{eq:rx antenna 2}, the resulting signal at the RX is
\begin{equation}
\begin{aligned}
\label{eq:y}
    \mathbf{y}(t)&=\mathbf{S}\mathbf{\tilde{y}}(t)=\mathbf{S}\sum_{l=0}^{L}\sum_{p=0}^{N_{\mathrm{sc}}-1}\sum_{q=0}^{M_{\mathrm{sym}}-1}\alpha_l\mathbf{r}(\theta_l)d_{pq}\\&e^{j2\pi p\Delta f(t-\tau_l-qT_{\mathrm{s}}-T_{\mathrm{cp}})}
    e^{j2\pi f_{D,l}t}\mathrm{rect}\left ( \frac{t-\tau_l-qT_{\mathrm{s}}}{T_{\mathrm{s}}}\right )+\mathbf{S}\mathbf{z}(t).
\end{aligned}
\end{equation}

After RSN, the received signal in \eqref{eq:y} is first divided into $M_{\mathrm{sym}}$ blocks with equal block duration $T_{\mathrm{s}}$. The $q$th block is given by:
\begin{equation}
    \mathbf{y}_q(t)=\mathbf{y}\left( t+qT_{\mathrm{s}} \right)\mathrm{rect}\left(\frac{t}{T_{\mathrm{s}}}\right).
\end{equation}

Then, the CP removal is performed on $\mathbf{y}_q(t)$:
\begin{equation}
    \bar{\mathbf{y}}_q(t)=\mathbf{y}_q(t+T_{\mathrm{cp}})\mathrm{rect}\left(\frac{t}{T}\right).
\end{equation}

Assuming that the CP duration is larger than the maximum delay, i.e., $T_{\mathrm{cp}}>\max(\tau_l)$, we have
\begin{equation}\label{eq:yq(t) before fft}
\begin{aligned}
\bar{\mathbf{y}}_q(t)&=\mathbf{S}\sum_{l=0}^{L}\sum_{p=0}^{N_{\mathrm{sc}}-1}\alpha_l\mathbf{r}(\theta_l)d_{pq}e^{j2\pi p\Delta f(t-\tau_l)}\\
&e^{j2\pi f_{D,l}(t+qT_{\mathrm{s}}+T_{\mathrm{cp}})}\mathrm{rect}\left ( \frac{t}{T}\right )
+\mathbf{S}\mathbf{z}(t+qT_{\mathrm{s}}+T_{\mathrm{cp}}).
\end{aligned}
\end{equation}

Within one symbol duration, the Doppler term $e^{j2\pi f_{D,l}(t+qT_{\mathrm{s}}+T_{\mathrm{cp}})}$ can be approximated as a constant $e^{j2\pi f_{D,l}(qT_{\mathrm{s}}+T_{\mathrm{cp}})}$. The waveform \eqref{eq:yq(t) before fft} is sampled at an interval of $T/N_{\mathrm{sc}}$, and then $N_{\mathrm{sc}}$-point fast Fourier transform (FFT) is performed on each symbol to obtain a spatial-frequency-time domain tensor $\mathcal{Y}\in\mathbb{C}^{N_{\mathrm{RF}}\times N_{\mathrm{sc}}\times M_{\mathrm{sym}}}$:
\begin{equation}
\label{eq:tensor y}
\mathcal{Y}_{:,p,q}=\mathbf{S}\sum_{l=0}^{L}\bar\alpha_l\mathbf{r}(\theta_l)d_{pq}e^{-j2\pi p\Delta f\tau_l}e^{j2\pi f_{D,l}qT_{\mathrm{s}}}+\mathbf{S}\mathbf{Z}_{p,q},
\end{equation}
where $\bar\alpha_l=\alpha_le^{j2\pi f_{D,l}T_{\mathrm{cp}}}$, $\mathbf{Z}_{p,q}=\frac{1}{N_{\mathrm{sc}}}\mathrm{FFT}\left\{ \mathbf{z}(iT/N_{\mathrm{sc}}+qT_{\mathrm{s}}+T_{\mathrm{cp}}) \right\}_i,1\le i\le N_{\mathrm{sc}}$. Suppose $z_{m,n}(t)$ follows i.i.d. CSCG distribution with variance $\sigma^2$ at each time sample, i.e., $z_{m,n}(t)\sim\mathcal{CN}(0,\sigma^2)$. Then we can get $\mathbf{z}(t)\sim\mathcal{CN}(0,M\sigma^2\mathbf{I}_N),\mathbf{Z}_{p,q}\sim\mathcal{CN}(0,\frac{M}{N_{\mathrm{sc}}}\sigma^2\mathbf{I}_{N})$.

\subsubsection{Communication Model}
From the communication perspective, the received signal \eqref{eq:tensor y} at the Rx can be written as
\begin{equation}
    \mathbf{y}_{pq}^{\mathrm{c}}=\mathbf{h}_{pq}^{\mathrm{c}}d_{pq}+\mathbf{n}_{pq}^{\mathrm{c}},
\end{equation}
where $\mathbf{h}_{pq}^{\mathrm{c}}=\mathbf{S}\sum_{l=0}^{L}\bar\alpha_l\mathbf{r}(\theta_l)e^{-j2\pi p\Delta f\tau_l}e^{j2\pi f_{D,l}qT_{\mathrm{s}}}$ is the equivalent communication channel for the $p$th subcarrier of OFDM symbol $q$, and $\mathbf{n}_{pq}^{\mathrm{c}}=\mathbf{S}\mathbf{Z}_{p,q}$ denotes the AWGN vector with each entry having zero mean and $\frac{M}{N_{\mathrm{sc}}}\sigma^2$ variance. Accordingly, the expected uplink communication rate is given by:
\begin{equation}
    \mathcal{R}=\frac{1}{BT_{\mathrm{s}}}\mathbb{E} \Bigg[ \frac{1}{M_{\mathrm{sym}}}\sum_{q=0}^{M_{\mathrm{sym}}-1}\sum_{p=0}^{N_{\mathrm{sc}}-1}\log_2\Big(1+\frac{\|\mathbf{h}_{pq}^{\mathrm{c}}\|^2P_t}{\frac{M}{N_{\mathrm{sc}}}\sigma^2}\Big) \Bigg],
\end{equation}
where $\mathbb{E}[\cdot]$ means takes expectation w.r.t $\alpha_l,\theta_l,\tau_l,f_{D,l}$.

\subsubsection{Sensing Model}
With the communication symbols $d_{pq}$ demouldated, we can perform data removal on \eqref{eq:tensor y}:
\begin{equation}
\label{eq:sense data}
\begin{aligned}
\bar{\mathcal{Y}}_{:,p,q}&=\frac{\mathcal{Y}_{:,p,q}}{d_{pq}}=\mathbf{S}\sum_{l=0}^{L}\bar\alpha_l\mathbf{r}(\theta_l)e^{-j2\pi p\Delta f\tau_l}e^{j2\pi f_{D,l}qT_{\mathrm{s}}}+\frac{\mathbf{S}\mathbf{Z}_{p,q}}{d_{pq}}\\
&=\sum_{l=0}^{L}\bar\alpha_l\mathbf{h}_{\mathrm{s}}(\theta_l)e^{j\omega_u^l p}e^{j\omega_v^l q}+\mathbf{n}_{s},
\end{aligned}
\end{equation}
where $\mathbf{h}_{\mathrm{s}}(\theta_l)=\mathbf{S}\mathbf{r}(\theta_l)$ is the equivalent steering vector, $\omega_u^l=-2\pi\Delta f\tau_l$ and $\omega_v^l=2\pi f_{D,l}T_{\mathrm{s}}$ are phase coefficient of subcarrier and symbol, respectively, and $\mathbf{n}_{s}=\frac{\mathbf{S}\mathbf{Z}_{p,q}}{d_{pq}}$ denotes the noise vector. Based on $\eqref{eq:sense data}$, we try to obtain $\theta_l,\tau_l$ and $f_{D,l}$. 




\subsection{Sensing Algorithm Design}
\subsubsection{AoA Estimation}
The spatial doamin of \eqref{eq:sense data} only contains angular information of targets $\theta_l$, so the subcarrier and symbol domain can be treated as snapshots. To this end, the $N_{\mathrm{RF}}\times N_{\mathrm{sc}}\times M_{\mathrm{sym}}$ dimensional signal tensor $\bar{\mathcal{Y}}$ in \eqref{eq:sense data} can be reorganized as a $N_{\mathrm{RF}}\times N_{\mathrm{sc}} M_{\mathrm{sym}}$ dimensional matrix $\mathbf{Y}_{\theta}$ as below
\begin{equation}\label{eq:tensor2matrix}
\mathbf{Y}_{\theta}(:,n_{p,q})=\bar{\mathcal{Y}}_{:,p,q},
\end{equation}
where $ n_{p,q}=p+qM_{\mathrm{sym}}$.
Thus, $\mathbf{X}_{\theta}$  can be equivalently expressed in matrix form as
\begin{equation}\label{eq:music snapshot}
    \mathbf{Y}_{\theta}=\mathbf{H}_\theta\mathbf{X}_\theta,
\end{equation}
where $\mathbf{H}_\theta\in\mathbb{C}^{N_{\mathrm{RF}}\times (L+1)}$ is the array manifold matrix, and $\mathbf{X}_\theta\in\mathbb{C}^{(L+1)\times N_{\mathrm{sc}}M_{\mathrm{sym}}}$, given by
\begin{equation}
\mathbf{H}_{\theta}=\Big[\mathbf{h}_{\mathrm{s}}(\theta_0),\mathbf{h}_{\mathrm{s}}(\theta_1),...,\mathbf{h}_{\mathrm{s}}(\theta_{L})\Big].
\end{equation}

\begin{equation}
\mathbf{X}_{\theta}(l,n_{p,q})=\bar\alpha_le^{j\omega_u^lp}e^{j\omega_v^lq}.
\end{equation}

The AoA estimation problem is to estimate the angles $\mathbb{S}_\theta = \{\theta_0,\theta_1,..., \theta_L\}$ of the $L+1$ localization/sensing targets embedded in the array manifold matrix $\mathbf{H}_\theta$. Note that different from the conventional arrays where the array manifold matrix are usually rotational invariance, the $\mathbf{H}_\theta$ for RAA no longer has such property. Therefore, we adopt the MUSIC algorithm which leverages the orthogonality between the noise subspace and the signal subspace to estimate target parameters. The required subspaces are derived through eigenvalue decomposition (EVD) of the covariance matrix for the observed data $\mathbf{Y}_{\theta}$ in \eqref{eq:music snapshot}, which can be expressed as
\begin{equation}\label{eq:covariance}
\begin{aligned}
    \mathbf{C}_{\mathbf{Y}}&=\frac{1}{N_{\mathrm{sc}}M_{\mathrm{sym}}}\mathbf{Y}_{\theta}^{\mathrm{H}}\mathbf{Y}_{\theta}\\
    &=\mathbf{E}_{\mathrm{s}}\mathbf{\Sigma}_{\mathrm{s}}\mathbf{E}_{\mathrm{s}}^\mathrm{H}+\mathbf{E}_{\mathrm{n}}\mathbf{\Sigma}_{\mathrm{n}}\mathbf{E}_{\mathrm{n}}^\mathrm{H},
\end{aligned}
\end{equation}
where $\mathbf{\Sigma}_{\mathrm{s}}\in\mathbb{C}^{(L+1)\times (L+1)},\mathbf{\Sigma}_{\mathrm{n}}\in\mathbb{C}^{(N_{\mathrm{RF}}-L-1)\times (N_{\mathrm{RF}}-L-1)}$ are diagonal matrices composed by the $L+1$ largest eigenvalues and the remaining $N_{\mathrm{RF}}-L-1$ eigenvalues, respectively. $\mathbf{E}_{\mathrm{s}}$ and $\mathbf{E}_{\mathrm{n}}$ denote the signal and noise subspaces, respectively. The MUSIC spatial spectrum is subsequently formulated as
\begin{equation}\label{eq:music specturm}
    P_{\mathrm{MUSIC}}(\theta)=\frac{1}{\mathbf{h}_{\mathrm{s}}^\mathrm{H}(\theta)\mathbf{E}_{\mathrm{n}}\mathbf{E}_{\mathrm{n}}^\mathrm{H}\mathbf{h}_{\mathrm{s}}(\theta_0)}.
\end{equation}

As such, the peaks of the MUSIC spectrum correspond to the estimation of AoAs of sensing targets. The estimation result is denoted by $\hat{\mathbb{S}}_\theta=\{\hat\theta_0,\hat\theta_1,..., \hat\theta_{L_k}\}$, where $L_k\le L+1$ due to the limitation of resolution or noise.

\subsubsection{Delay and Doppler Estimation}

\begin{algorithm}[!h]
    \caption{Proposed Algorithm for OFDM-ISAC with RAA}
    \label{alg:algorithm}
    \renewcommand{\algorithmicrequire}{\textbf{Input:}}
    \renewcommand{\algorithmicensure}{\textbf{Output:}}
    \begin{algorithmic}[1]
        \REQUIRE Signal tensor $\bar{\mathcal{Y}}$ in \eqref{eq:sense data}   
        \ENSURE $\theta_{l}$, $\tau_l$, $f_{D,l}$, $\forall l$    
        
        \STATE Reorganize $\bar{\mathcal{Y}}_{:,p,q}$ as $\mathbf{Y}_\theta$ in \eqref{eq:tensor2matrix}

        \STATE Calculate the covariance matrix $\mathbf{C}_{\mathbf{Y}}$ of $\mathbf{Y}_\theta$ with \eqref{eq:covariance}

        \STATE Perform EVD of the covariance of $\mathbf{C}_{\mathbf{Y}}$, and obtain the noise subspace $\mathbf{E}_{\mathrm{n}}$

        \FOR{ $\theta \in [-\theta_\mathrm{max},\theta_\mathrm{max}]  $ }
            \STATE Calculate the equivalent steering vector $\mathbf{h}_{\mathrm{s}}(\theta)$ in \eqref{eq:sense data}
            \STATE Calculate the searching spectrum $P_{\mathrm{MUSIC}}(\theta)$ in \eqref{eq:music specturm}
            
        \ENDFOR
        \STATE Find the $L_k\le L+1$ highest peaks of $P_{\mathrm{MUSIC}}(\theta)$ to obtain the estimated AoAs $\hat{\mathbb{S}}_\theta=\{\hat\theta_0,\hat\theta_1,..., \hat\theta_{L_k}\}$

        \FOR{ $l \in \{0,1,...,L_k\}$}
            \STATE Calculate ZF beamforming vector $\mathbf{h}_{\mathrm{ZF}}(\hat{\theta_{l}})$ in \eqref{eq:spatial beamforming}
            \STATE Perform spatial filtering on $\bar{\mathcal{Y}}_{:,p,q}$ in \eqref{eq:target k matrix} to obtain $\mathbf{Y}^l_{p,q}$
            \STATE Perform Periodogram algorithm on $\mathbf{Y}^l_{p,q}$ in \eqref{eq:periodogram} to obtain $\bar{\mathbf{Y}}^l_{p',q'}$
            \STATE Search the peak of $\bar{\mathbf{Y}}^l_{p',q'}$ and transform them into desired parameters $\hat\tau_l,\hat f_{D,l}$ in \eqref{eq:delay Doppler est}

        \ENDFOR

        \STATE Match the parameters and obtain $\{ \theta_l,\tau_l, f_{D,l} \}$

        \RETURN $ \theta_l,\tau_l, f_{D,l} , l=\{ 0,1,...,L_k\}$
    \end{algorithmic}
\end{algorithm}

Owing to the super-resolution capability of the MUSIC algorithm, closely spaced AoAs can be effectively discriminated, which enables sequential parameter estimation through spatial beamforming. Furthermore, the signal tensor $\bar{\mathcal{Y}}_{:,p,q}$ in \eqref{eq:sense data} exhibits distinctive structural properties across its subcarrier and symbol dimensions: The signal phase increases linearly w.r.t. subcarrier index $p$ and symbol index $q$. This structural regularity constitutes critical prior knowledge that can be exploited to simplify the estimation problem and reduce computational complexity. For range and Doppler estimation, we therefore implement a two-stage approach:
\begin{itemize}
    \item ZF spatial beamforming to isolate target-specific signals.
    \item Periodogram algorithm leveraging the spectral characteristics in subcarrier and symbol domains.
\end{itemize}

This methodology takes advantage of the OFDM waveform structure while maintaining estimation accuracy and computational efficiency through decoupled processing of spatial and spectral parameters.

Specially, the ZF beamforming vectors are given by
\begin{equation}\label{eq:spatial beamforming}
\mathbf{h}_{\mathrm{ZF}}(\hat{\theta_{l}})=\frac{\hat{\mathbf{H}}_l\mathbf{h}_{\mathrm{s}}(\hat\theta_{l})}{\left \| \hat{\mathbf{H}}_l\mathbf{h}_{\mathrm{s}}(\hat\theta_{l}) \right \|_2},
\end{equation}
where 
\begin{equation}\label{eq:hat H}
    \hat{\mathbf{H}}_l=\mathbf{I}_M-\mathbf{H}_l(\mathbf{H}_l^{\mathrm{H}}\mathbf{H}_l)^{-1}\mathbf{H}_l^{\mathrm{H}}.
\end{equation}

\begin{equation}\label{eq:H_l}
    \mathbf{H}_l=\begin{bmatrix}
\mathbf{h}_{\mathrm{s}}(\hat\theta_0)   & ... & \mathbf{h}_{\mathrm{s}}(\hat\theta_{l-1})&\mathbf{h}_{\mathrm{s}}(\hat\theta_{l+1})&...&\mathbf{h}_{\mathrm{s}}(\hat\theta_{L-1})
\end{bmatrix}.
\end{equation}

Then for target $k$, the beamformed signal matrix $\mathbf{Y}^k\in \mathbb{C}^{N_{\mathrm{sc}}\times M_{\mathrm{sym}}}$ can be obtained as
\begin{equation}\label{eq:target k matrix}
\begin{aligned}
\mathbf{Y}^k_{p,q}&=\mathbf{h}_{\mathrm{ZF}}^{\mathrm{H}}(\hat{\theta}_{k})\bar{\mathcal{Y}}_{:,p,q}\\
&=\mathbf{h}_{\mathrm{ZF}}^{\mathrm{H}}(\hat{\theta}_{k})\sum_{l=0}^{L}\bar\alpha_l\mathbf{h}_{\mathrm{s}}(\theta_l)e^{j\omega_u^lp}e^{j\omega_v^lq}+\mathbf{h}_{\mathrm{ZF}}^{\mathrm{H}}(\hat{\theta}_{k})\mathbf{n}_{s}\\
&\stackrel{(a)}{\approx}\mathbf{h}_{\mathrm{ZF}}^{\mathrm{H}}(\hat{\theta}_{k})\mathbf{h}_{\mathrm{s}}(\theta_k)e^{j\omega_u^lp}e^{j\omega_v^lq}+\mathbf{h}_{\mathrm{ZF}}^{\mathrm{H}}(\hat{\theta}_{k})\mathbf{n}_{s},
\end{aligned}
\end{equation}
where $(a)$ holds because $\mathbf{h}_{\mathrm{ZF}}^{\mathrm{H}}(\hat{\theta}_{k})\mathbf{h}_{\mathrm{s}}(\theta_l)\approx0,k\ne l$.

By substituting $\omega_u^l=-2\pi\Delta f\tau_l$ and $\omega_v^l=2\pi f_{D,l}T_{\mathrm{s}}$ into \eqref{eq:target k matrix}, we consider $(N_{\mathrm{sc}}, M_{\mathrm{sym}})$-point 2D-Periodogram algorithm on $\mathbf{Y}^l_{p,q}$ 
\begin{equation}\label{eq:periodogram}
\begin{aligned}
&\bar{\mathbf{Y}}^l_{p',q'}= \mathrm{FFT}\bigg\{  \mathrm{IFFT}\Big\{\mathbf{Y}^l_{p,q}\Big\}_p \bigg\}_q  \\
&=\bar\alpha_l\mathbf{h}_{\mathrm{ZF}}^{\mathrm{H}}(\hat{\theta_{l}})\mathbf{h}_{\mathrm{s}}(\theta_l)
e^{j\pi(N_{\mathrm{sc}}-1)(\frac{p'}{N_{\mathrm{sc}}}-\Delta f\tau_l)}\\
&\quad\quad \times e^{j\pi(M_{\mathrm{sym}}-1)(\frac{q'}{ M_{\mathrm{sym}}}-T_{\mathrm{s}} f_{D,l})} \frac{\sin\big(\pi N_{\mathrm{sc}}(\frac{p'}{N_{\mathrm{sc}}}-\Delta f\tau_l)\big)}{\sin\big(\pi (\frac{p'}{N_{\mathrm{sc}}}-\Delta f\tau_l)\big)}\\
&\quad\quad\times\frac{\sin\big(\pi M_{\mathrm{sym}}(\frac{q'}{ M_{\mathrm{sym}}}-T_{\mathrm{s}} f_{D,l})\big)}{\sin\big(\pi (\frac{q'}{ M_{\mathrm{sym}}}-T_{\mathrm{s}}f_{D,l})\big)}\\
&=\alpha_{lp'q'}\mathbf{h}_{\mathrm{ZF}}^{\mathrm{H}}(\hat{\theta_{l}})\mathbf{h}_{\mathrm{s}}(\theta_l)\frac{\sin\big(\pi N_{\mathrm{sc}}(\frac{p'}{N_{\mathrm{sc}}}-\Delta f\tau_l)\big)}{\sin\big(\pi (\frac{p'}{N_{\mathrm{sc}}}-\Delta f\tau_l)\big)}\\
&\quad\quad\times\frac{\sin\big(\pi M_{\mathrm{sym}}(\frac{q'}{ M_{\mathrm{sym}}}-T_{\mathrm{s}} f_{D,l})\big)}{\sin\big(\pi (\frac{q'}{ M_{\mathrm{sym}}}-T_{\mathrm{s}}f_{D,l})\big)},
\end{aligned}
\end{equation}
where $\alpha_{lp'q'}$ denotes the equivalent path coefficient, and $\mathrm{FFT}\{\cdot\}_p,\mathrm{IFFT}\{\cdot\}_q$ means perform FFT and inverse FFT (IFFT) algorithm on a designated dimension, respectively. Therefore, we have obtained the spectrum of delay and Doppler of target $l$. By searching the peaks of 2-D matrix $|\bar{\mathbf{Y}}^l_{p',q'}|^2$, those indexes $\hat{p}'_l,\hat{q}'_l$ of peak values can be transformed into our desired parameters:
\begin{align}\label{eq:delay Doppler est}
\hat\tau_l=\frac{\hat{p}'_l}{N_{\mathrm{sc}}\Delta f},\quad \hat f_{D,l}=\frac{\hat{q}'_l}{M_{\mathrm{sym}}T_{\mathrm{s}}},
\end{align}
where $\hat\tau_l$ and $\hat f_{D,l}$ are the estimated delay and Doppler of target $l$, respectively. For target $l$, all the parameters $\{ \theta_l,\tau_l, f_{D,l} \}$ can thus be obtained. Note that the resolution  limit of Periodogram algorithm depends on FFT/IFFT points, while the computational complexity also increases with it. Therefore, a proper value for the size of the FFT/IFFT should be selected. 

The pseudo-code of the proposed sensing algorithm is summarized in Algorithm~\ref{alg:algorithm}.

\section{Simulation Results}\label{sec:simulation}
In this section, we provide numerical results to verify the performance of RAA in ISAC system. For RAA, we set the maximum orientation of sULAs and elements in each sULA as $\eta_{\max}=\pi/2,M=128$ at Rx, respectively. Therefore, the total number of sULAs needed is $N=2\left \lfloor \frac{\eta_{\mathrm{max}}}{\arcsin(2/M)}+1 \right \rfloor=201$. The orientation of each sULA is $\eta_n = n\arcsin(2/M),n\in\mathcal{N}$, and the distance $D$ is $\lambda/4(\sin(0.5\arcsin(2/M)))$. For HBF, the number of DFT codewords is $N'=\frac{2}{2/M}=128$, and the $n$th codeword’s angle is $\varphi_n=\arcsin(\frac{n-9}{8}),1\le n\le N'$. The number of RF chains is set to $N_{\mathrm{RF}}=8$. The carrier frequency is set as $f_c=39$GHz, the subcarrier spacing as $\Delta f=120$kHz, and the number of subcarriers, and symbols in one CPI as $N_{\mathrm{sc}}=512,M_{\mathrm{sym}}=2048$, respectively. The system bandwidth as be obtained as $B=N_{\mathrm{sc}}\Delta f=61.44$MHz. 
In addition, the radiation patterns of antenna elements follow the 3GPP antenna model, which is expressed in dB as \cite{3GPPchannel}
\begin{equation}
    G_{\mathrm{dB}}(\theta)=G_0^{\mathrm{dB}}-\min\{ 12(\theta/\theta_{\mathrm{3dB}})^2,A_{\max}^{\mathrm{dB}} \},
\end{equation}
where $G_0^{\mathrm{dB}}$ is the peak antenna gain in dB, $A_{\max}^{\mathrm{dB}}=30\mathrm{dB}$ denotes the front-to-back attenuation and $\theta_{\mathrm{3dB}}$ accounts for the 3dB beamwidth. For ULA, $\theta_{\mathrm{3dB}}^{\mathrm{ULA}}$ is set to $\pi$ to cover the entire direction, while for RAA, $\theta_{\mathrm{3dB}}^{\mathrm{RAA}}$ is set to $0.3\pi$, as each sULA is only responsible for a narrower AoA range. Besides, $G_0^{\mathrm{dB}}$ in ULA and RAA are set to 0dB and 5.1335dB to guarantee the same total power gain for all directions. The parameter settings are listed in Tab. \ref{tab:parameter}

\begin{figure}[t]
  \centering
  \subfloat[\label{fig:MUSIC RAA m}]{
    \includegraphics[width=0.9\linewidth]{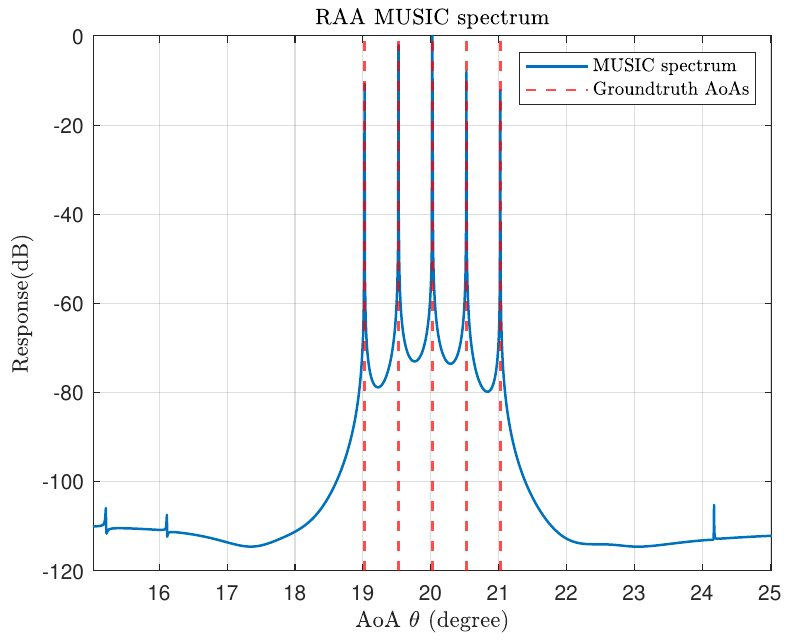}
  }

  \subfloat[\label{fig:MUSIC HBF m}]
  {
    \includegraphics[width=0.9\linewidth]{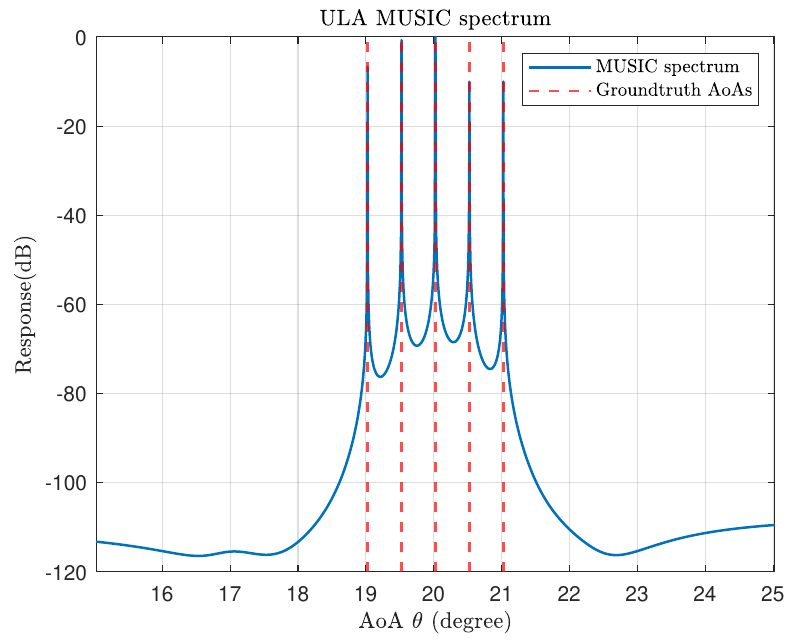}
  }
  \caption{MUSIC specturm of (a) RAA (b) ULA with moderate AoAs}
  \label{fig:MUSIC m}
\end{figure}

\begin{table}[H]
\caption{\textbf{System Settings}}\label{tab:parameter}
\centering
\begin{tabular}{ccc}
\toprule
Parameter & Symbol & Value  \\
\midrule
Carrier frequency & $f_c$ & 39GHz \\
System bandwidth & $B$ & 61.44MHz \\
Number of subcarriers & $N_{\mathrm{sc}}$ & 512 \\
Number of OFDM symbols & $M_{\mathrm{sym}}$ & 2048 \\
Subcarrier spacing & $\Delta f$ & 120kHz \\
OFDM symbol duration & $T$ & 8.33$\mu$s \\
CP duration & $T_{\mathrm{cp}}$ & 0.67$\mu$s \\
Total symbol duration & $T_{\mathrm{s}}$ & 9$\mu$s \\
Number of antenna elements in sULA/ULA & $M$ & 128 \\
Number of sULAs & $N$ & 201 \\
Number of RF-chains & $N_{\mathrm{RF}}$ & 8 \\
Number of DFT codewords & $N'$ & 128 \\
Ratio of transmit power and noise& $P_t/\sigma^2$ & 20dB \\
\bottomrule
\end{tabular}
\end{table}


\begin{figure}[t]
  \centering
  \subfloat[\label{fig:MUSIC RAA l}]{
    \includegraphics[width=0.9\linewidth]{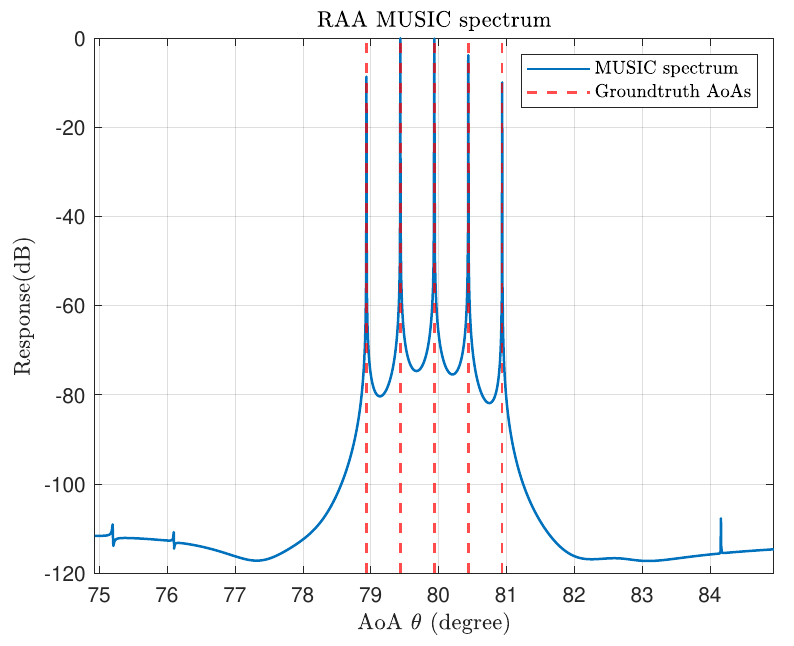}
  }

  \subfloat[\label{fig:MUSIC HBF l}]
  {
    \includegraphics[width=0.9\linewidth]{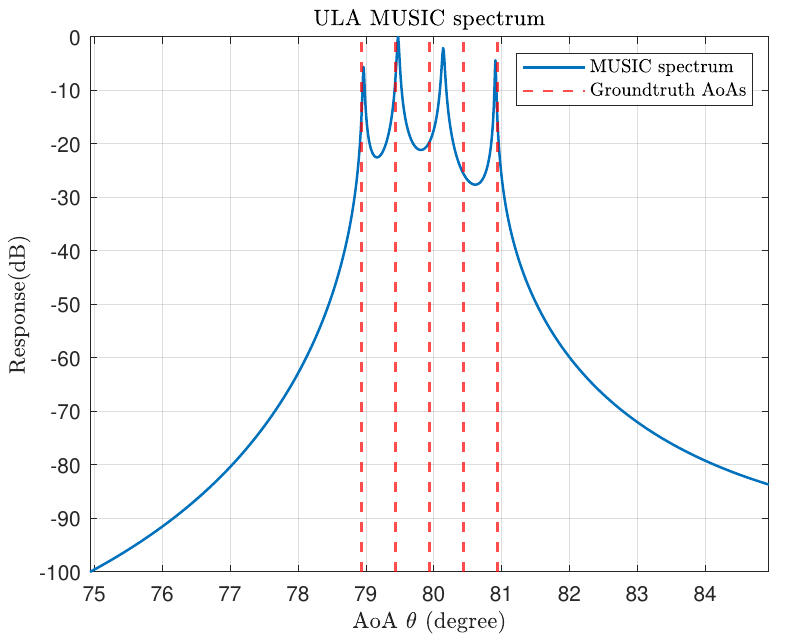}
  }
  \caption{MUSIC specturm of (a) RAA (b) ULA with large AoAs}
  \label{fig:MUSIC l}
\end{figure}

The UAV swarm comprises five low-altitude UAVs arranged with equal 0.5-degree angular spacing in their AoAs. By deliberately adjusting the geometric centroid of the swarm, distinct mean AoA values can be achieved for different sensing scenarios. The delay and Doppler frequency parameters of the targets are set following the Gaussian distribution, where the mean of delay is in accordance with the adjusted swarm centroid, the mean of Doppler is 300Hz, and their variances are $4\times10^{-16}$s$^2$ and $6400$Hz$^2$, respectively. Both RAA and ULA architectures are considered to illustrate the sensing 
parameter estimation performance, which is evaluated in terms of the RMSE of the estimated angle

Note that due to the close proximity of UAV swarms, adjacent targets may not be perfectly discriminated, so another criterion, termed as \textit{average missing shots}, is incorporated to evaluate the performance, which is defined as
\begin{equation}
    \varepsilon  =\frac{1}{Q}\sum_{i=1}^{Q}\Big( \mathrm{card}\big(\mathbb{S}_\theta^i \big)-\mathrm{card}\big(\hat{\mathbb{S}}_\theta^i \big) \Big),
\end{equation}
where $Q$ denotes the total number of testing rounds, $\mathbb{S}_\theta^i$ and $\hat{\mathbb{S}}_\theta^i$ denote the real and estimation angle set of $i$th test round, respectively. We assume that $\mathrm{card}\big(\mathbb{S}_\theta^i \big)\ge\mathrm{card}\big(\hat{\mathbb{S}}_\theta^i \big)$, which holds when the noise is rather small.





Fig. \ref{fig:MUSIC m} and Fig. \ref{fig:MUSIC l} demonstrate the MUSIC specturm of AoA estimation with RAA and ULA under large and moderate AoAs. It can be seen that when targets are around the array boresight, i.e., with relatively small AoAs, both RAA and ULA can distinguish them very well. However, if the AoAs are large, ULA fails to discriminate the different UAVs in the swarm, while RAA still have an excellent performance. This is consistent with the the resolution comparison between ULA and RAA in Fig. \ref{fig:resolution compare}.

\begin{figure*}[t]
  \centering
  \subfloat[\label{fig:RD map for all}]{
    \includegraphics[width=0.32\linewidth]{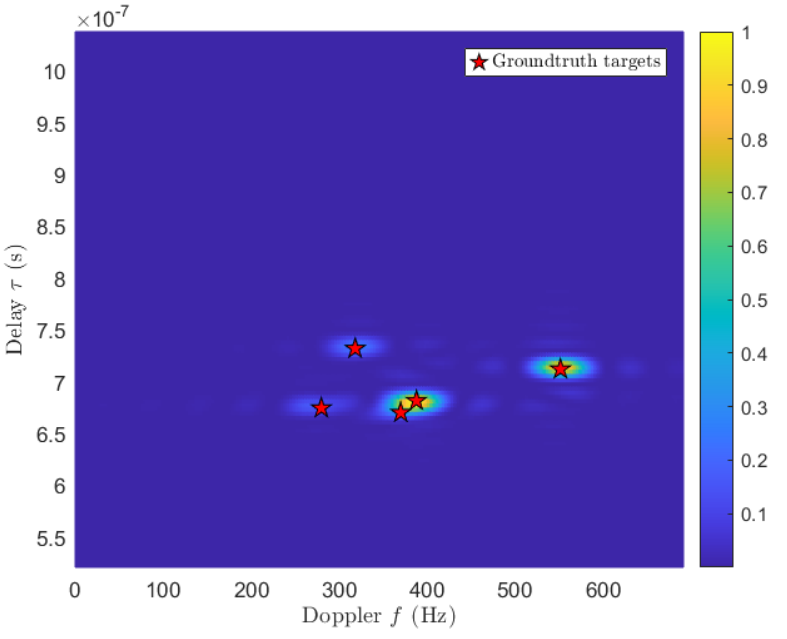}
  }
  \subfloat[\label{fig:RD map 1}]
  {
    \includegraphics[width=0.32\linewidth]{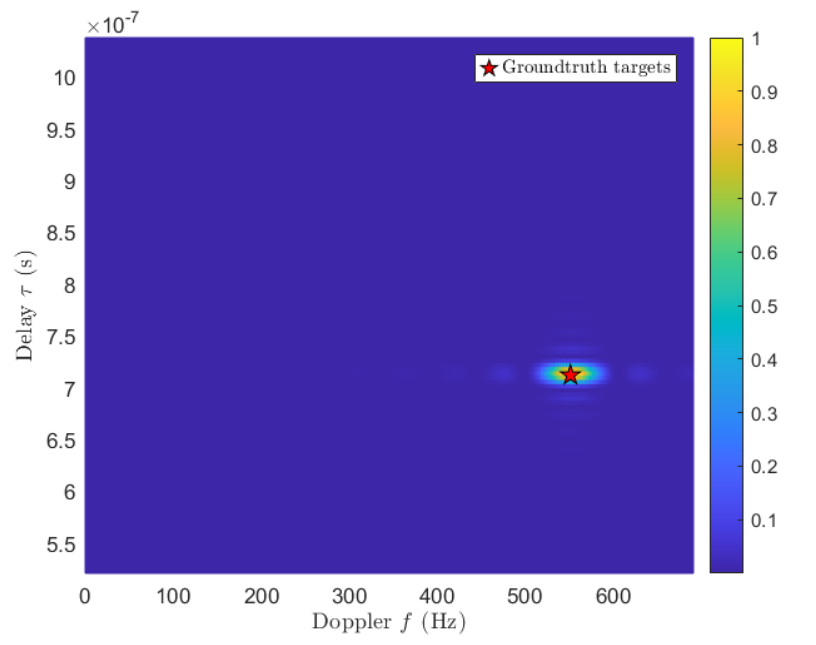}
  }
  \subfloat[\label{fig:RD map 2}]
  {
    \includegraphics[width=0.32\linewidth]{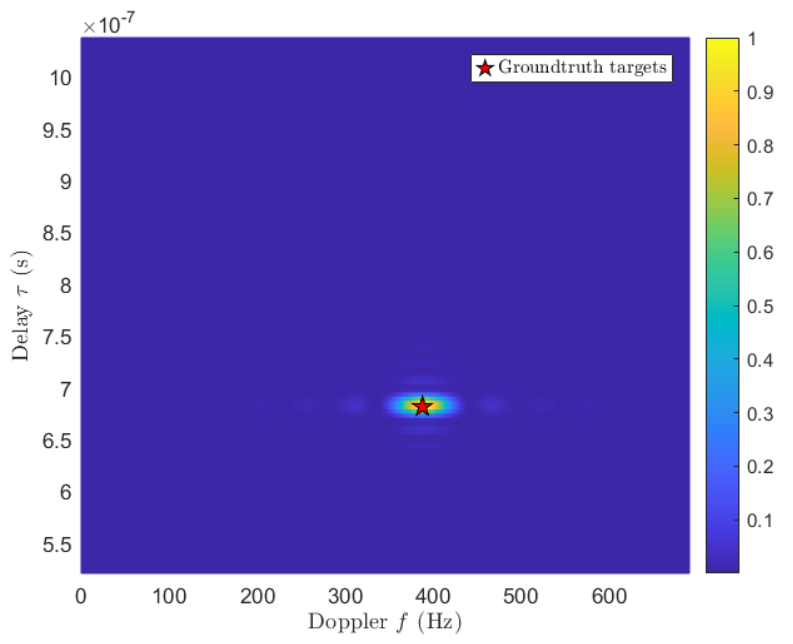}
  }

  \subfloat[\label{fig:RD map 3}]
  {
    \includegraphics[width=0.32\linewidth]{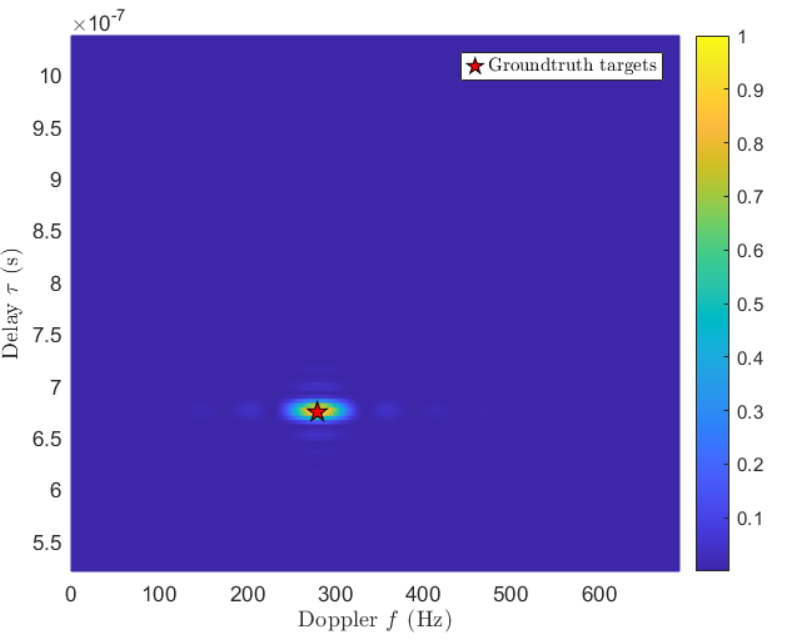}
  }
  \subfloat[\label{fig:RD map 4}]
  {
    \includegraphics[width=0.32\linewidth]{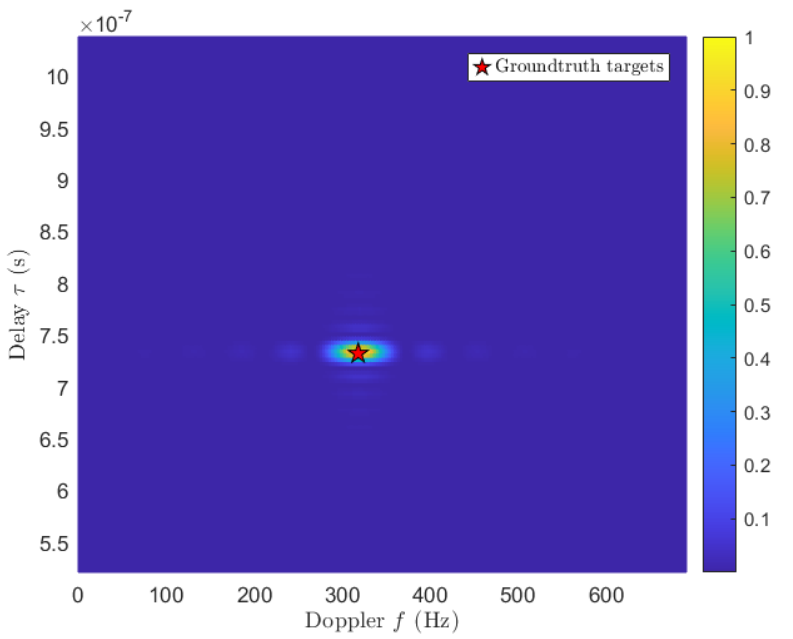}
  }
  \subfloat[\label{fig:RD map 5}]
  {
    \includegraphics[width=0.32\linewidth]{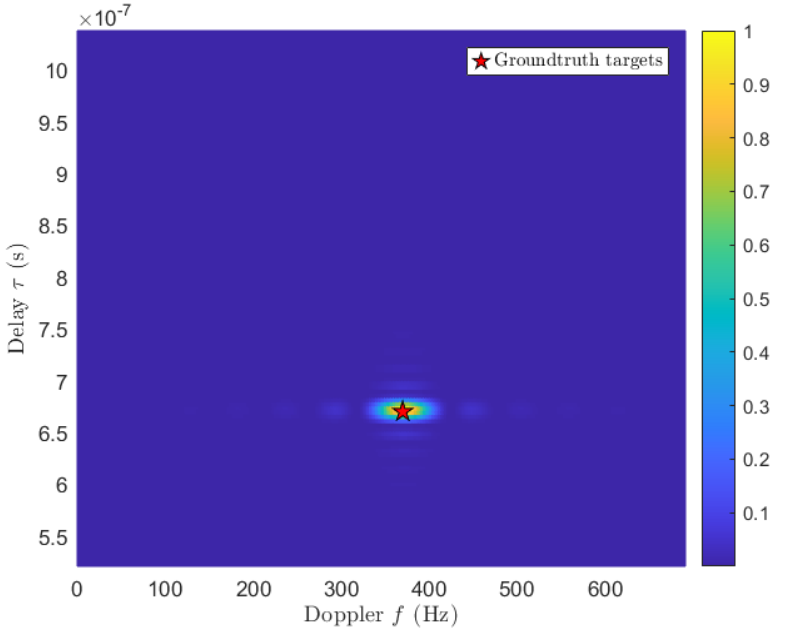}
  }
  \caption{An example of Delay-Doppler maps of targets (a) without ZF beamforming, (b)-(f) with ZF beamforming}
  \label{fig: RD map}
\end{figure*}

Fig. \ref{fig:RD map for all} demonstrates the delay and Doppler information about all the five targets. It can be inferred that if spatial ZF beamforming is not performed, some of the targets may be buried in the noise, or overlapping with each other making them difficult to distinguish. With ZF beamforming, the delay and Doppler information of each target are shown in Fig. \ref{fig:RD map 1}-\ref{fig:RD map 5}, respectively. Owing to the super resolution ability of MUSIC algorithm, all targets can be filtered out by ZF beamforming and thus achieves excellent estimation accuracy in delay and Doppler by oversampling and zero padding in Periodogram algorithm.

\begin{figure}[t] 
        \centering \includegraphics[width=0.9\columnwidth]{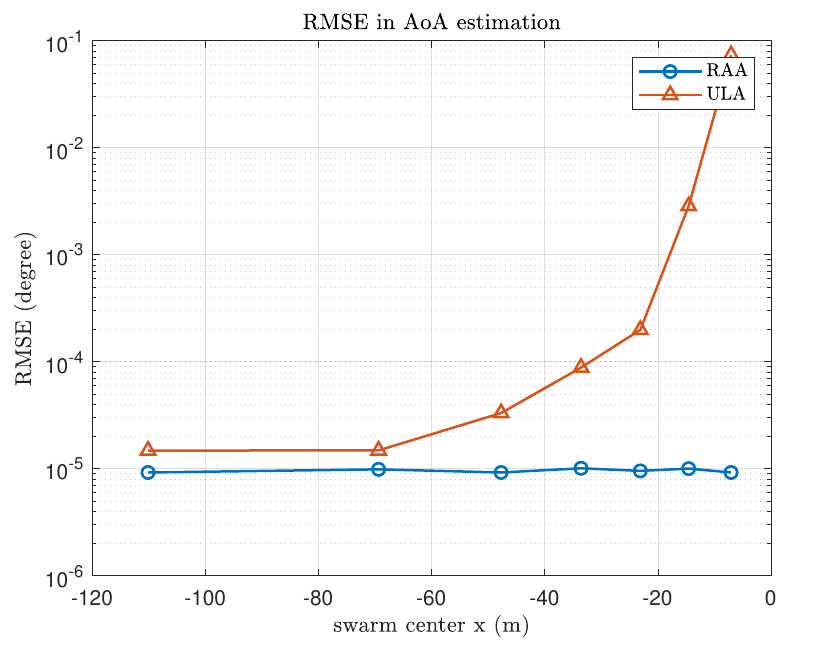}
        \caption{\label{fig:AOArmse}RMSE of AoA estimation}
\end{figure}

\begin{figure}[t] 
        \centering \includegraphics[width=0.9\columnwidth]{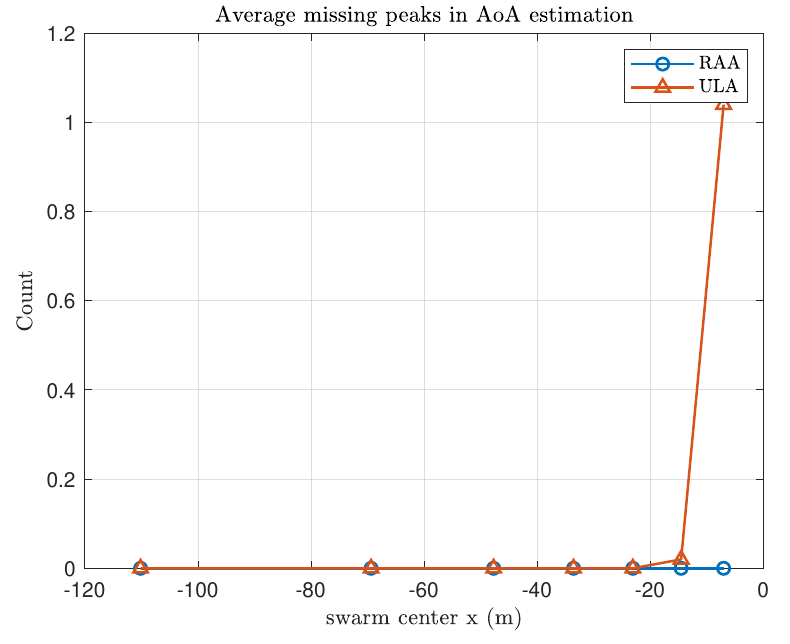}
        \caption{\label{fig:missing peaks}Average missing shots of AoA estimation}
\end{figure}

Fig. \ref{fig:AOArmse} shows the comparison of AoA estimation RMSE between RAA and ULA w.r.t. different AoAs centroid. It can be seen that as the UAV swarm moving closer with increasing AoAs, the RMSE of ULA increases dramatically while that of RAA remains almost the same. This can be explained that as the resolution of ULA degrades, adjacent AoAs cannot be discriminate and they form fake peaks, resulting in error in AoA estimation. Besides, the limited resolution also brings about missing shots, which is obvious when AoAs are large, as shown in Fig. \ref{fig:missing peaks}.

\begin{figure}[t] 
        \centering \includegraphics[width=0.9\columnwidth]{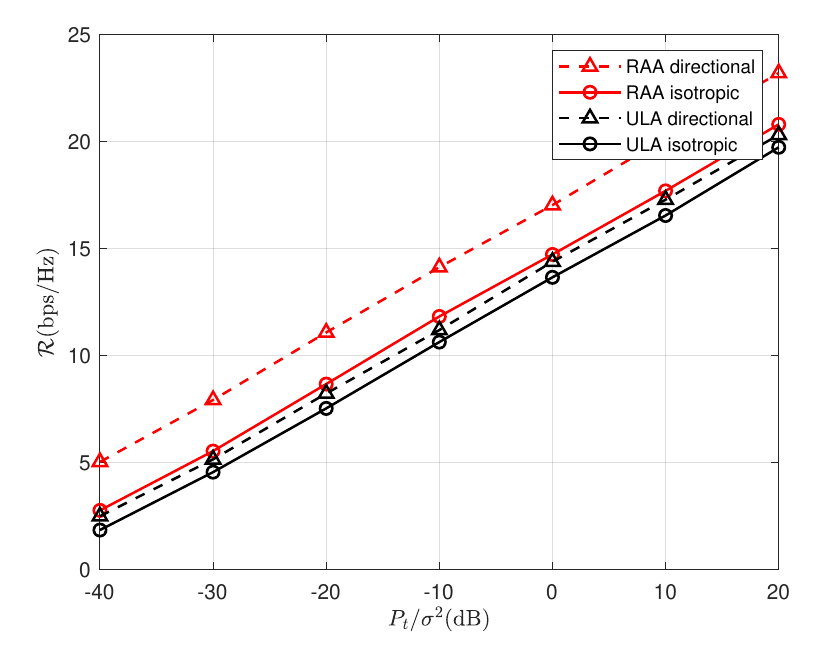}
        \caption{\label{fig:communication rate}Communication rate for RAA and ULA.}
\end{figure}







Fig. \ref{fig:communication rate} plots the achievable communication rate of RAA and ULA. The isotropic antenna element gain in any direction is set as $G_{\mathrm{dB}}^{\mathrm{iso}}=-2.816$dB to guarantee the same total power gain for all directions. It can be observed that RAA outperforms ULA in terms of communication performance as well, especially when directional antenna elements are exploited. This can be attributed to the higher directional antenna gain since each sULA is responsible for a narrower range of incoming signals. These results demonstrate that RAA can achieve better communication and sensing performance compared to ULA.


\section{Conclusion}\label{sec:conclusion}

In this paper, we proposed and investigated the RAA-based ISAC systems for low-altitude UAV swarm. By analyzing the beam pattern of RAA, we rigorously showed that unlike conventional ULAs, RAA can achieve uniform angular resolution across all signal directions. We also demonstrated that RAA can achieve higher overall beamforming gain than the conventional ULA since antenna elements with higher directivity can be used. Besides, we proposed efficient sensing algorithms for RAA based UAV swarm ISAC with OFDM. MUSIC-based AoA estimation, ZF spatial filtering, and 2-D Periodogram enabled robust multi-target AoA-delay-Doppler detection under OFDM signaling. Numerical results demonstrated the effectiveness of the proposed RAA based ISAC system, which outperforms conventional ULA in both target estimation and communication performance.


{\appendices
\section{Proof of Theorem 3}\label{app:resolution}
Based on Theorem \ref{the:RAA resolution} and \ref{the:HBF resolution}, in $\gamma_{\mathrm{ULA}}(\theta')$ we denote $\sin\theta'$ by $x$, and take derivative of $\gamma_{\mathrm{ULA}}(\arcsin(x))$ w.r.t. $x$, we will get:
\begin{equation}
\begin{aligned}
\frac{\mathrm{d} \gamma_{\mathrm{ULA}}}{\mathrm{d} x} &=\frac{1}{2}\frac{\mathrm{d} \big[ \arcsin(x+\frac{2}{M})-\arcsin(x-\frac{2}{M})\big]}{\mathrm{d} x} \\
&=\frac{1}{2}\Bigg( \frac{1}{\sqrt{1-(x+\frac{2}{M})^2}}-\frac{1}{\sqrt{1-(x-\frac{2}{M})^2}} \Bigg) \\
&=\frac{1}{2}\frac{\sqrt{1-(x-\frac{2}{M})^2}-\sqrt{1-(x+\frac{2}{M})^2}}{\sqrt{1-(x+\frac{2}{M})^2}\sqrt{1-(x-\frac{2}{M})^2}}\\
&=\frac{1}{2}\frac{(x+\frac{2}{M})^2-(x-\frac{2}{M})^2}
{\Delta}\\
&=\frac{4x}{M\Delta}\\
\frac{\mathrm{d}^2 \gamma_{\mathrm{ULA}}}{\mathrm{d} x^2}&=\frac{4}{M\Delta}
\end{aligned}
\end{equation}


where $\Delta=\sqrt{1-u^2}\sqrt{1-v^2}(\sqrt{1-v^2}+\sqrt{1-u^2})\ge0,u=x+\frac{2}{M},v=x-\frac{2}{M},$ and $\frac{\mathrm{d}^2 \gamma_{\mathrm{ULA}}}{\mathrm{d} x^2}\ge0$. Therefore, $x=0$ is the global minimum of $\gamma_{\mathrm{ULA}}(\arcsin(x))$ and $\gamma_{\mathrm{ULA}}(\theta')=\gamma_{\mathrm{ULA}}(\arcsin(x))\ge\gamma_{\mathrm{ULA}}(0)=\arcsin(\frac{2}{M})=\gamma_{\mathrm{RAA}}$.

 }

\bibliographystyle{IEEEtran}      
\bibliography{IEEEabrv,reference}

\end{document}